\documentclass[a4paper,fleqn,usenatbib]{mnras}
\usepackage{array, multirow, bigdelim, makecell, booktabs} 
\usepackage[T1]{fontenc}
\usepackage{ae,aecompl}
\usepackage{graphicx}	% Including figure files
\usepackage{amsmath}	% Advanced maths commands
\usepackage{amssymb}	% Extra maths symbols
\usepackage{topcapt}
\usepackage{supertabular}
\usepackage{adjustbox}
\usepackage{hyperref}
\usepackage{stfloats}
\hypersetup{
    colorlinks=true,
    linkcolor=blue,
    filecolor=magenta,      
    urlcolor=cyan,
}
\usepackage{newtxtext,newtxmath}

\title[Galaxy / cluster co-evolution]{An extreme case of galaxy and cluster co-evolution at \emph{z}\,=\,0.7}
\author[Ebeling et al.]{
H.\ Ebeling$^1$\thanks{E-mail: ebeling@ifa.hawaii.edu}, J.\ Richard$^2$, I.\ Smail$^3$, A.\,C.\ Edge$^3$, A.\ M.\ Koekemoer$^4$, L.\ Zalesky$^1$
\\ \\
$^1$ Institute for Astronomy University of Hawaii, 2680 Woodlawn Drive Honolulu, HI 96822, USA\\
$^2$ Univ Lyon, Univ Lyon1, Ens de Lyon, CNRS, Centre de Recherche Astrophysique de Lyon UMR5574, F-69230, Saint-Genis-Laval, France\\
$^3$ Centre for Extragalactic Astronomy, Department of Physics, Durham University, South Road, Durham DH1 3LE, UK\\
$^4$ Space Telescope Science Institute, 3700 San Martin Dr., Baltimore, MD 21218, USA
}
\date{Accepted 2021 September 16. Received 2021 September 2; in original form 2020 December 2}
\pubyear{2020}
\begin{document}
\renewcommand{\floatpagefraction}{1}

\label{firstpage}
\pagerange{\pageref{firstpage}--\pageref{lastpage}}
\maketitle

\begin{abstract}
We report the discovery of eMACS\,J0252.4$-$2100 (eMACS\,J0252), a massive and highly evolved galaxy cluster at $z=0.703$. Our analysis of {\it Hubble Space Telescope} imaging and VLT/MUSE and Keck/DEIMOS spectroscopy of the system finds a high velocity dispersion of 1020$^{+180}_{-190}$\,km\,s$^{-1}$ and a high (if tentative) X-ray luminosity of  $(1.2\pm 0.4)\times10^{45}$\,erg\,s$^{-1}$ (0.1--2.4 keV). As extreme is the system's brightest cluster galaxy, a giant cD galaxy that forms stars at a rate of between 85 and 300\,M$_\odot$\,yr$^{-1}$ and features an extended halo of diffuse [\ion{O}{II}] emission, as well as evidence of dust. Its most remarkable properties, however, are an exceptionally high ellipticity and a radially symmetric flow of gas in the surrounding intracluster medium, potential direct kinematic evidence of a cooling flow. A strong-lensing analysis, anchored by two multiple-image systems with spectroscopic redshifts, finds the best lens model to consist of a single cluster-scale halo with a total mass of $(1.9\pm0.1)\times 10^{14}$\,M$_\odot$ within 250\,kpc of the cluster core and, again, an extraordinarily high ellipticity of $e=0.8$. Although further, in-depth studies across the electromagnetic spectrum (especially in the X-ray regime) are needed to conclusively determine the dynamical state of the system, the properties established so far suggest that eMACS\,J0252 must have already been highly evolved well before $z\sim 1$, making it a prime target to constrain the physical mechanisms and history of the co-evolution or dark-matter halos and baryons in the era of cluster formation.
\end{abstract}


\section{Introduction}
\label{sec:intro}

%
% Figure 1
%
\begin{figure*}
\includegraphics[width=0.98\textwidth]{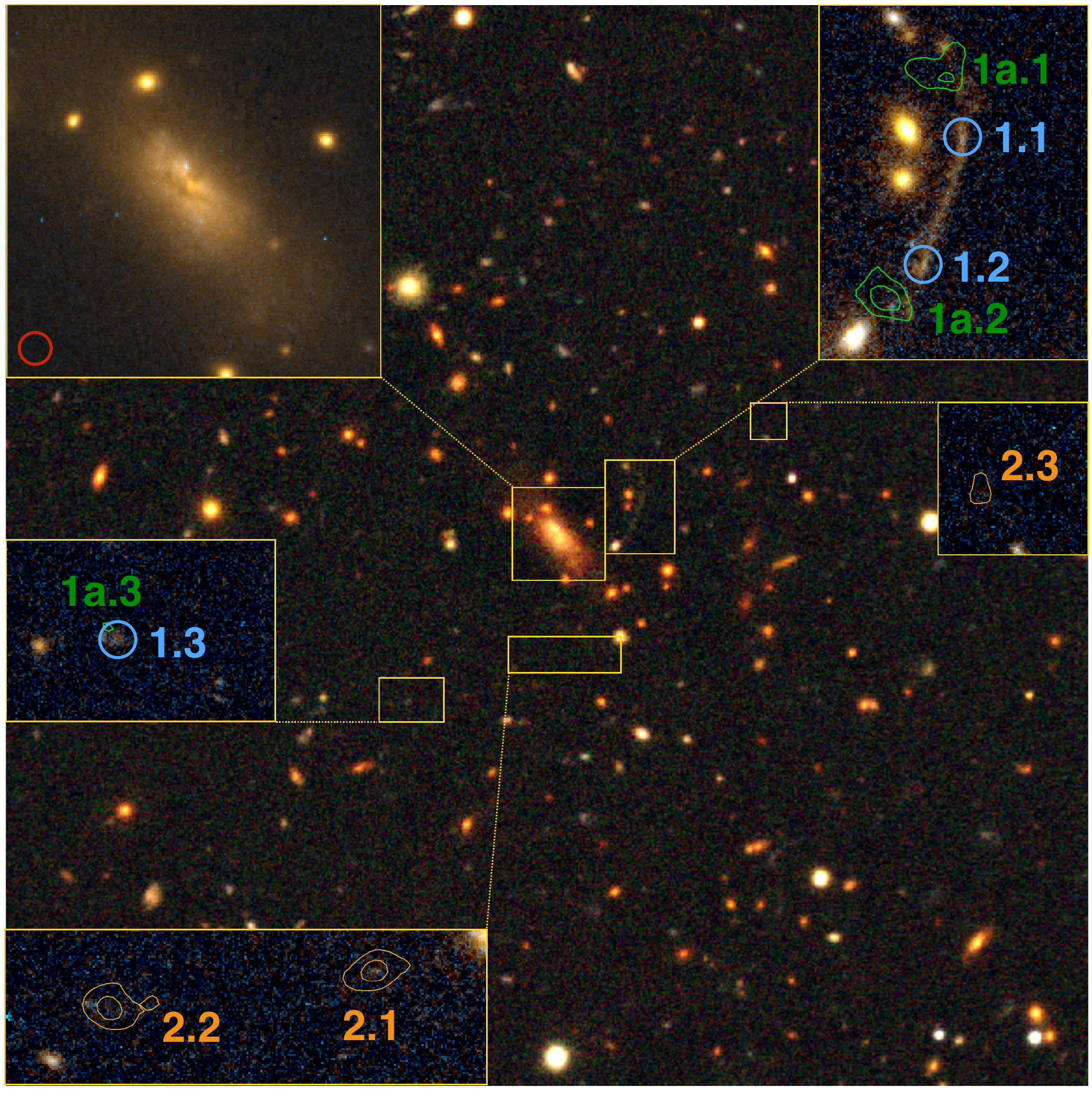}
\caption{Groundbased image ($g$, $r$, $i$) of eMACS\,J0252 as obtained with Gemini-N/GMOS, covering 1 Mpc on the side at the cluster redshift. The insets show the BCG as well as (at increased contrast) strong-lensing features as seen with {\it HST}/WFC3  (F200LP+F110W; GO-15703) and MUSE (Ly$\alpha$ contours, see Section~\ref{sec:SL} for details). For illustrative purposes, the red circle shows the approximate size of the seeing disk of our groundbased observations. Note the very extented stellar distribution in the BCG and signs of both a dust lane and dual nuclei. 
\label{fig:image}}
\end{figure*}

In the widely accepted paradigm of hierarchical structure formation, galaxy clusters form through successive mergers. As a result, the fraction of undisturbed (``relaxed") clusters decreases monotonically with increasing redshift in a trend that becomes profound at $z>0.4$ \citep{2012MNRAS.420.2120M} but may weaken again at yet higher redshifts \citep{2017ApJ...841....5N}. In parallel, the giant elliptical galaxies at the hearts of clusters grow through similar merger and accretion processes, resulting in brightest cluster galaxies (BCGs) that represent the by far most massive concentrations of stars in the Universe. It has long been known that the evolution of BCGs and their host clusters are closely linked \citep[e.g.,][]{1984ARA&A..22..185D,1989ARA&A..27..235K,2004ApJ...617..879L,2007MNRAS.375....2D,2009AJ....137.4795C,2014ApJ...797...82L}; however, to which extent the rate of BCG evolution tracks that of the host cluster remains a subject of debate \citep[][and references therein]{2016MNRAS.460.2862B}.

cD galaxies constitute a special class of BCG: found almost exclusively in fully relaxed clusters, they are characterized most readily by their enormous sizes, extreme luminosities, and (in rich clusters) extended, low-surface-brightness halos \citep{1964ApJ...140...35M,1976ApJ...209..693O,1986ApJS...60..603S,2013ApJ...765...25N,2013MNRAS.436..275W}. Not only are such extreme galaxies absent from the low-density field; even within the already overdense environment of galaxy clusters they invariably mark the center of pronounced mass overdensities \citep[e.g.,][]{1983ApJ...274..491B}. In recognition of the unique position of cD galaxies at the extreme end of the galaxy mass and luminosity functions, BCG evolution and its interconnection with the cluster environment remains an area of intense study. Of particular interest are the questions of whether BCG growth occurs mostly early on \citep[suggested by numerical simulations and supported by the fact that cD galaxies can form already in high-density galaxy groups;][]{1977ApJ...211..309A,1981ApJ...248..439T,1984ApJ...276...26M}; at intermediate times in the process of cluster mergers \citep[consistent with the high frequency of cDs with multiple nuclei;][]{1983ApJ...268..476S,1985AJ.....90.2431T,2020ApJS..247...43K}; or late, during cluster virialization \citep[supported, e.g., by the  prevalence of dust and star formation in cD galaxies in cool-core clusters, or the strong link between the luminous envelopes of cD galaxies and the distribution of intra-cluster light;][]{1989AJ.....98.2018M,2008ApJ...681.1035O,2020ApJS..247...43K}. 

If we are to understand which  physical mechanisms dominate the 
co-evolution of clusters and BCGs, we need to identify and study extreme cases across the widest accessible range of redshifts. The most striking example of a fully relaxed cluster hosting a BCG with enhanced star formation and nuclear activity is the Phoenix Cluster at $z=0.6$ \citep{2012Natur.488..349M,2014ApJ...784...18M}. At even higher redshifts, recent studies of distant optically and infra-red selected clusters have found evidence for enhanced star formation and AGN activity in BCGs at $z>1$ \citep{2017MNRAS.469.1259B, 2019MNRAS.487.1210T}. However, these clusters are not only significantly less massive than the majority of distant clusters selected by X-ray and SZ techniques, they are also not dynamically relaxed and do not host giant central galaxies like the Phoenix Cluster or, at lower redshift, A\,1835 \citep[$z=0.252$;][]{2006ApJ...648..164M}.

 Recent observations of eMACS\,J0252.4$-$2100 ($z=0.703$), the subject of this paper, offer an opportunity to obtain evidence of the interplay of BCG and cluster evolution in a redshift-mass regime poorly sampled by earlier work. Our paper is structured as follows: we  provide an overview of all observations used in our analysis in Section~\ref{sec:obs}, and discuss physical properties of the BCG and the host cluster in Section~\ref{sec:results}, before presenting our conclusions and plans for further study in Section~\ref{sec:summary}.

We assume the $\Lambda$CDM concordance cosmology with $\Omega_{m} = 0.3, \Omega_{\Lambda} = 0.7$, and $H_{0} = 100 h$ km s$^{-1}$ Mpc$^{-1}$, with $h = 0.7$.

%
% Figure 2
%
\begin{figure*}
    \centering
    \includegraphics[height=2.55in]{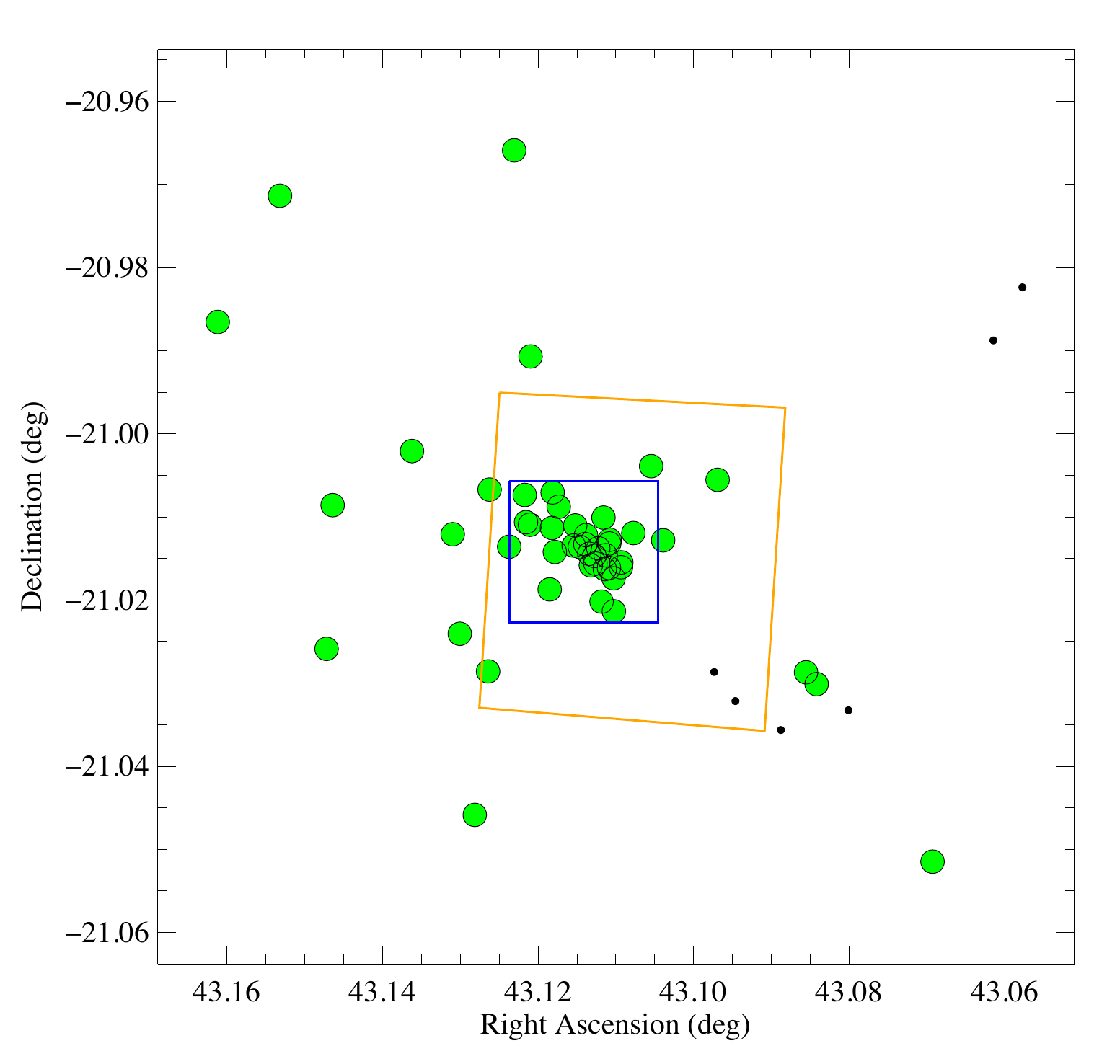}\includegraphics[height=2.6in]{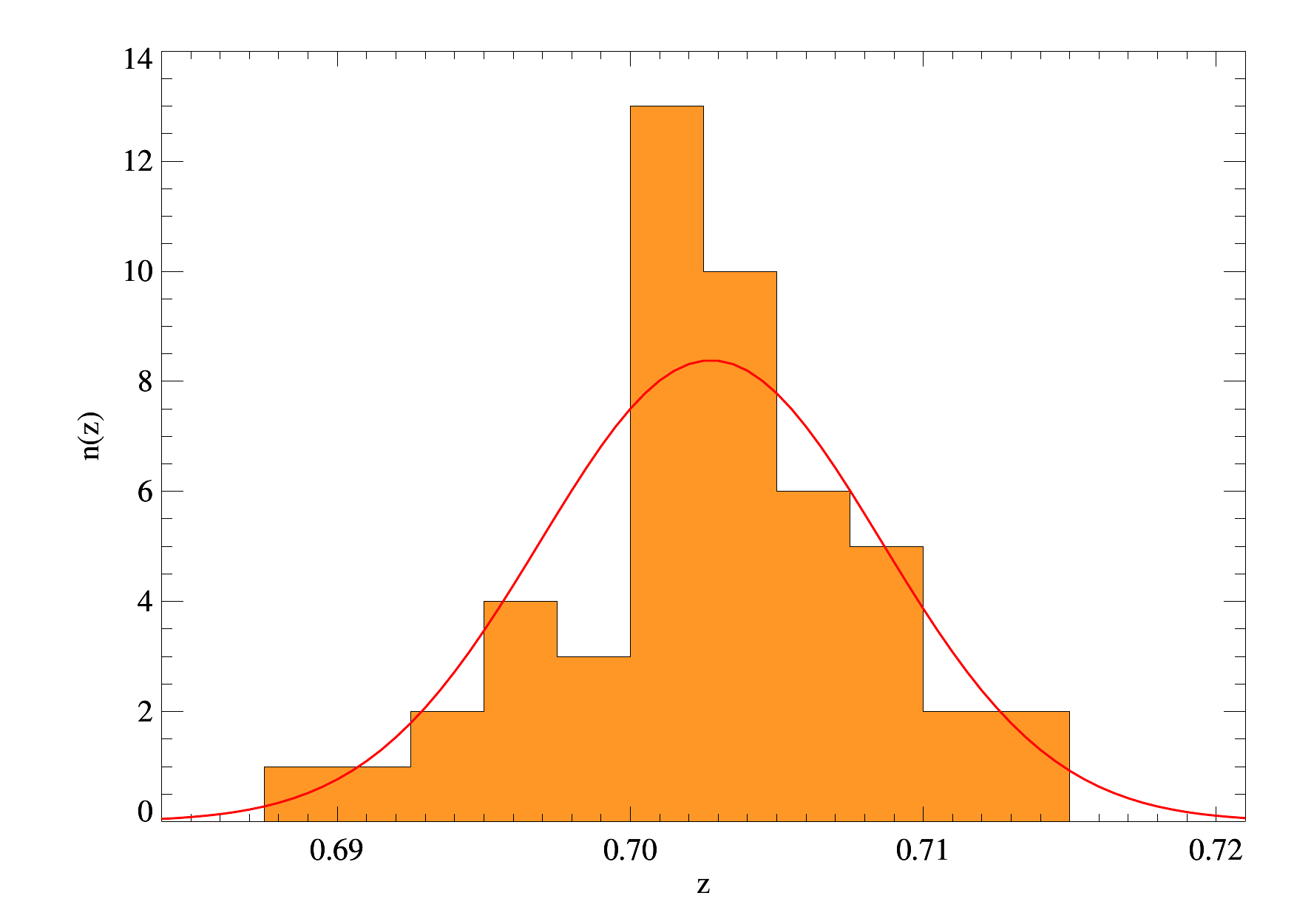}
    \caption{\textit{Left:} Spatial distribution of galaxies with spectroscopic redshifts in the range from $z$=0.65 to 0.75 from our DEIMOS and MUSE observations. Cluster members (redshifts within three times the cluster velocity dispersion) are highlighted as filled green circles; the blue and orange rectangles delineate the field of view of MUSE and WFC3/NIR, respectively. 
    Note the alignment of the distribution of cluster members with the major axis of the BCG from Figure~\ref{fig:image}.
    \textit{Right:} Histogram of the redshifts of spectroscopic galaxy members for eMACS\,J0252, as derived from the observations described in Section~\ref{sec:obs-spec}. The velocity dispersion of the overlaid, best-fit Gaussian is 1020$^{+180}_{-190}$ km\,s$^{-1}$. We find no strong evidence of substructure along the line of sight.
    }
    \label{fig:zsky}
\end{figure*}

\section{Observations and data reduction}
\label{sec:obs}

\subsection{eMACS\,J0252.4$-$2100}
\label{sec:target}

The galaxy cluster eMACS\,J0252.4$-$2100 ($z=0.703$, hereafter eMACS\,J0252) was discovered by the eMACS project \citep{2013MNRAS.432...62E} in November 2014, when spectroscopic follow-up observations of an overdensity of galaxies around the X-ray source 1RXS\,J025227.2$-$210031, detected in the {\it ROSAT} All-Sky Survey \citep[RASS;][]{1999A&A...349..389V}, established an approximate redshift of $z=0.698$ for the associated galaxy cluster. While the Sunyaev-Zel'dovich (SZ) signal of the system was missed by the {\it Planck} mission \citep[no corresponding entry is listed in the sample published in ][]{2016A&A...594A..27P}, the cluster was -- according to \citet{2019hst..prop16017G} -- detected in SZ observations conducted with the South Pole Telescope (SPT) and is listed as SPT-CLJ0252$-$2100 in the online supplement to \citet{2020ApJS..247...25B}.

\subsection{Optical and Near-infrared Imaging}
\label{sec:obs-im}

We observed eMACS\,J0252 with the GMOS camera on the Gemini-North 8.1-m telescope atop Maunakea on
September 9, 2015 (Program ID GN\,--2015B-Q-42, PI Ebeling). Three dithered exposures of 90s duration each were taken in each of the $g^\prime$, $r^\prime$, and $i^\prime$ passbands. We used standard data reduction procedures as provided by the Gemini data analysis pipeline to create bias-subtracted, flat-fielded, and fringe-corrected co-added images for each filter. A color image of a 1 Mpc square region around the cluster, based on our GMOS data, is shown in Fig.~\ref{fig:image}.

On Dec.\ 13, 2018, eMACS\,J0252 was observed with the Wide Field Camera 3 (WFC3) aboard the {\it Hubble Space Telescope} (\textit{HST}) for the SNAPshot program GO-15307 (PI Gladders). WFC3 obtained images in the F200LP and the F110W filters, using the camera's UVIS and IR channels, respectively. The former filter covers the UV and optical regime from approximately 2000\AA\ to 8000\,\AA, whereas the latter extends from 9000\,\AA\ to 1.4\,$\mu$m at near-infrared wavelengths. Three dithered  exposures were obtained in each filter, with total exposure times of 741s and 758s for the F200LP and F110W images, respectively. After initial default calibration by the {\it HST} archive pipeline, these images were subsequently processed through additional calibration and alignment routines, following approaches described in more detail in \citet{2011ApJS..197...36K}. In particular, these subsequent steps resulted in improved astrometric alignment between all the exposures, across both filters, to an accuracy of a few milliarcseconds (rms), as well as improved cleaning of cosmic rays and other detector defects compared to the original archival products. The final combined mosaics for both filters are drizzled to a scale of 30 milliarcseconds per pixel.

Images of salient features within the cluster core, created from the \textit{HST} data, are shown as insets in Fig.~\ref{fig:image}.

\subsection{Spectroscopy}
\label{sec:obs-spec}

After the initial identification of eMACS\,J0252 as an X-ray luminous galaxy cluster through redshift measurements with the DEIMOS spectrograph \citep{2003SPIE.4841.1657F} on the Keck-II 10m telescope on Maunakea (see Sec.~\ref{sec:target}), additional spectroscopic observations of presumed cluster member galaxies were performed with the same telescope and instrument in December 2018, using  again the 600 line mm$^{-1}$ grating and the GG455 blocking filter. Applying standard data reduction techniques, we extracted two- and one-dimensional spectra and determined redshifts from cross correlations with template spectra using an adapted version of the software package developed by \citet{2011PASP..123..638M}.

eMACS\,J0252 was subsequently observed with the MUSE Integral-Field Unit on the 8.2-m UT4 telescope of the European Southern Observatory's Very Large Telescope on Cerro Paranal (Chile) on Sep.\ 6, 2019. Three dithered 970s integrations were obtained in 0.95$^{\prime\prime}$ seeing for program 0103.A-0777(A) (PI Edge) in WFM-NOAO-N mode. The MUSE data were reduced using v2.7 of the MUSE data reduction pipeline \citep{2020A&A...641A..28W}, following standard reduction recipes for instrumental calibration and sky subtraction, including specific improvements for self-calibration and masking of interstacks as described in detail in \citet{2020arXiv200909784R}.
The field of view of the final combined datacube covers 1$\times$1 arcmin$^2$ centered on the cluster BCG, sampled at a 0.2$^{\prime\prime}$ pixel scale. The wavelength range covers 4750--9350\AA\ at 1.25\AA\ per pixel, with a spectral resolution $R\sim2700$. We extracted spectra from continuum sources using the \textit{HST}/F110W image as an input catalog. We also searched for line emitters in the reduced datacube using the {\tt muselet} software, which is part of the MPDAF python package \citep{2019ASPC..521..545P}.

The distribution on the equatorial sky of galaxies with spectroscopic redshifts in the range from $z=0.65$ to 0.75 obtained from our observations is shown in Fig.~\ref{fig:zsky}.

\subsubsection{Radio and Submillimetre}
\label{sec:radio}

A search of public radio surveys (TGSS, GLEAM, WISH, NVSS, and VLASS) at the location of the BCG of eMACS\,J0252 reveals a relatively bright, compact radio source detected from 76\,MHz to 3,GHz. 

A short SCUBA2 observation of eMACS\,J0252 was obtained at 850\,$\mu$m in August 2015 as part of a larger survey of eMACS clusters (PI Edge). The exposure of 50 min in good conditions ($\tau_{225}=0.06$) achieved an rms sensitivity of 4.2\,mJy per beam; the lack of a detection in this observation places 3-$\sigma$ upper limits of 12.6\,mJy on the BCG flux density
at 850\,$\mu$m.

\section{Analysis and Results}
\label{sec:results}

As is evident from Fig.~\ref{fig:image}, the optical appearance of eMACS\,J0252 is heavily dominated by the system's highly elliptical BCG. We here examine key physical properties of both the exceptional galaxy at the core of eMACS\,J0252 and of its cluster environment.

%
% Figure 3
%
\begin{figure*}
    \centering
    \includegraphics[width=0.48\textwidth]{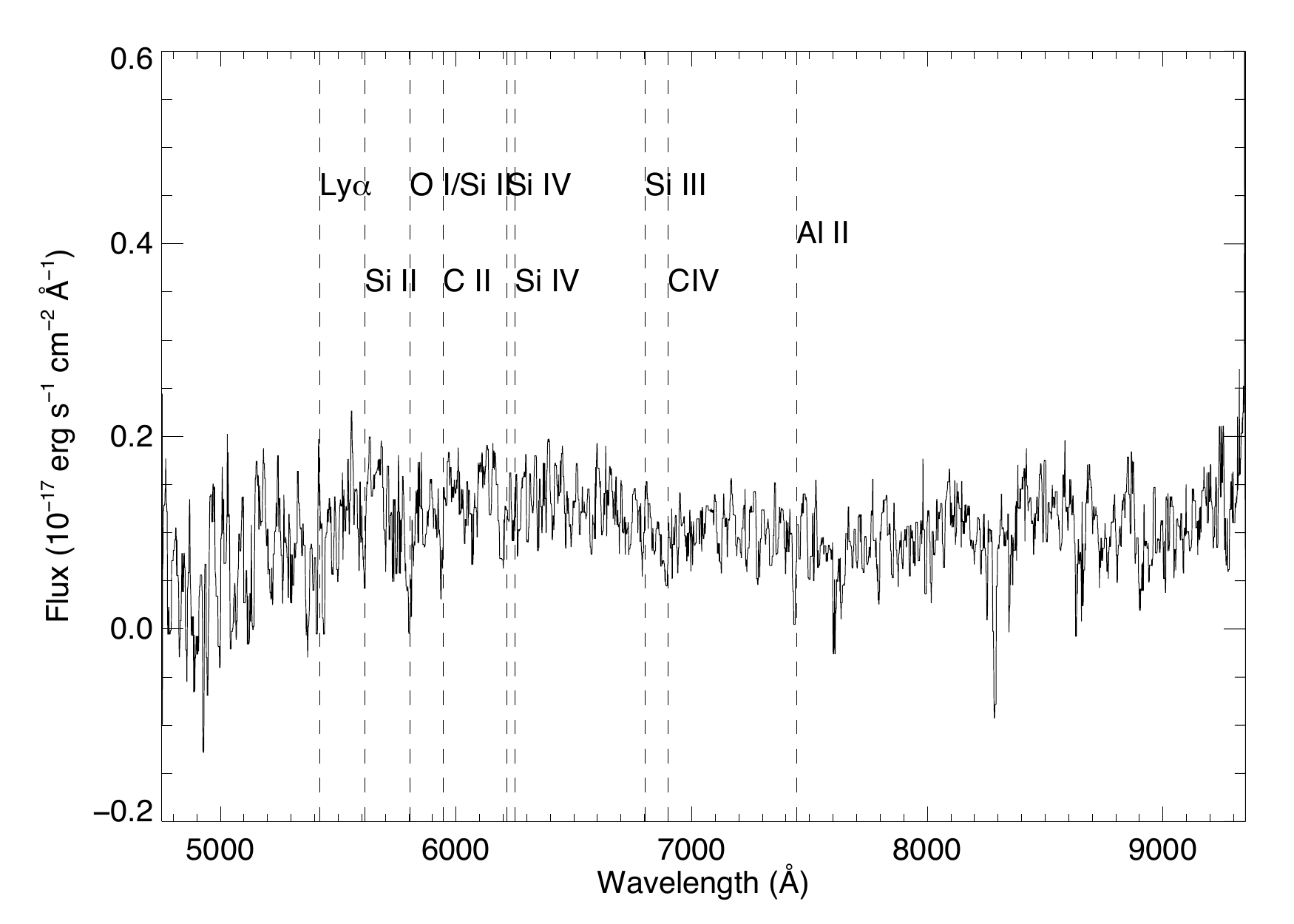}    \includegraphics[width=0.48\textwidth]{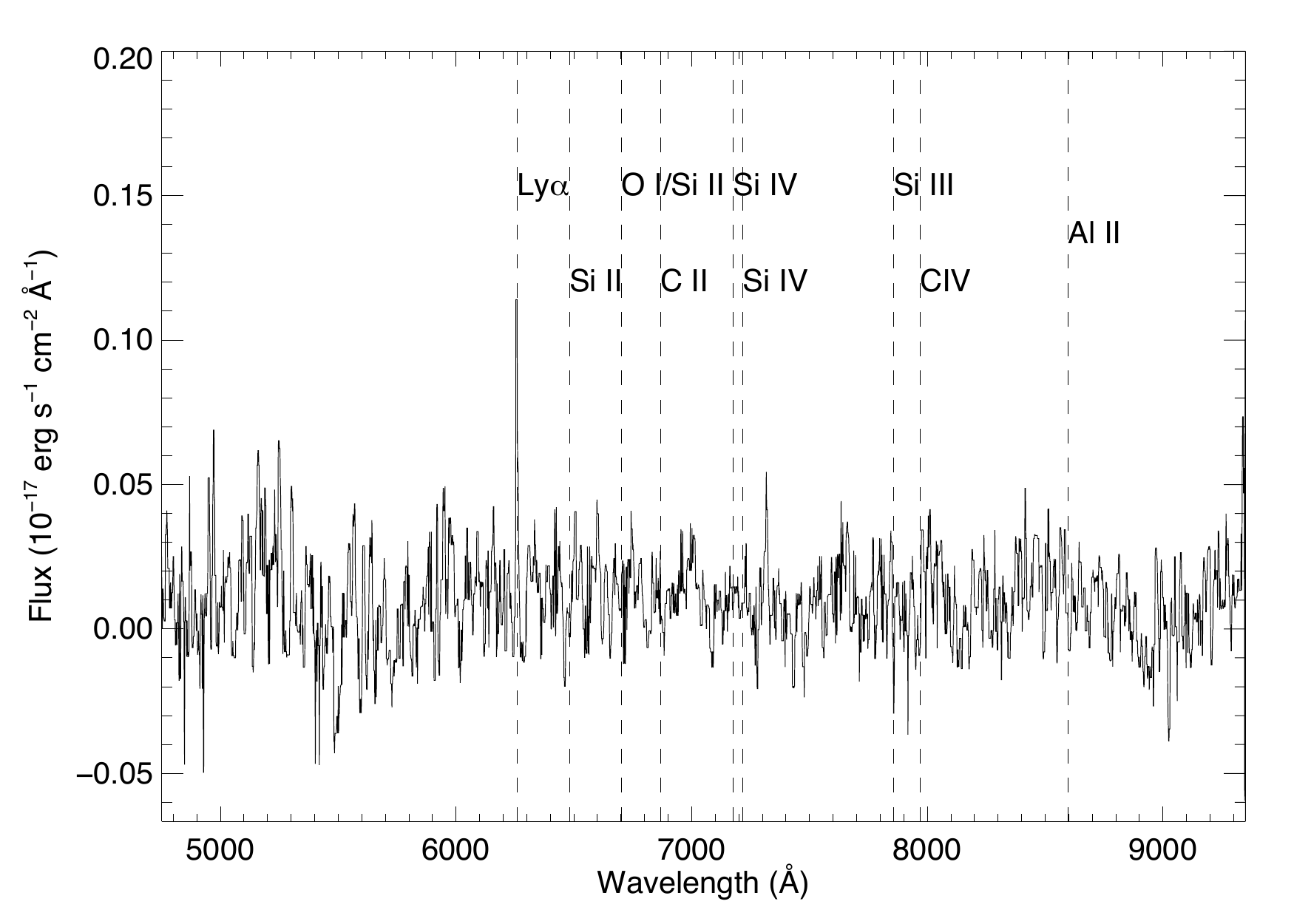}
    \caption{MUSE spectra of the two multiple-image systems identified by us in eMACS\,J0252. \textit{Left:} both components of System 1 ($z=3.458$); \textit{right:} Image 1 of System 2 ($z=4.148$).}
    \label{fig:SLspec}
\end{figure*}

\subsection{Cluster properties}

\subsubsection{Radial velocities}

The spectroscopic observations  described in Section~\ref{sec:obs-spec} yield a heliocentric cluster redshift of $z=0.7028$ and a velocity dispersion of $\sigma=1020^{+180}_{-190}$ km s$^{-1}$, based on the 49 concordant redshifts shown in Fig.~\ref{fig:zsky} and listed in the Appendix. 

We find no significant substructure along the line of sight in the overall redshift distribution (Fig.~\ref{fig:zsky}), as would be expected if the observed velocity dispersion were inflated by ongoing or recent merger activity perpendicular to the plane of the sky. 

The large clustercentric distances reached in our DEIMOS observations enable us to probe the large-scale environment of eMACS\,J0252 (Fig.~\ref{fig:zsky}) and test whether the system is isolated in three dimensions or connected to mass concentrations in the vicinity. However, the redshift distribution of the galaxies observed to date exhibits no clear trends with either right ascension or declination.

\subsubsection{X-ray properties}
At the cluster redshift of $z=0.703$, the X-ray flux detected in the RASS corresponds to an X-ray luminosity of $(1.2\pm 0.4)\times10^{45}$\,erg\,s$^{-1}$ (0.1--2.4 keV). Although, at face value, this measurement suggests that eMACS\,J0252 is one of a mere handful of exceptionally X-ray luminous clusters known at $z>0.6$, this assessment remains subject to significant uncertainty, owing to the RASS detection in the direction of the source consisting of a mere 12 net photons. A robust determination of the X-ray properties of eMACS\,J0252 will have to account for contamination from X-ray point sources, a task that requires much better angular resolution than is achieved in past or current X-ray all-sky surveys ({\it ROSAT} and {\it eROSITA}).

\subsubsection{Strong gravitational lensing}
\label{sec:SL}

Scrutiny of the spectra obtained with MUSE (Section~\ref{sec:obs-spec}) reveals unambiguous signs of strong gravitational lensing in the form of two multiple-image systems. The giant arc almost due West of the BCG is found to consist of two merged images of a galaxy at $z=3.458$. We show the MUSE spectrum of the arc in Fig.~\ref{fig:SLspec} (left); features anchoring the redshift measurement are faint Ly$\alpha$ emission spatially offset from the continuum (and included as subsystem 1a in our modeling of the mass distribution), as well as \ion{Si}{II} and \ion{C}{II} absorption. A second multiple-image system is spectroscopically identified through pronounced Ly$\alpha$ emission at 5420\AA, revealing the source redshift to be $z=4.148$ (Fig.~\ref{fig:SLspec}, right). The components of all multiple-image systems are listed in Table~\ref{tab:SL} and shown and labeled in the insets to Fig.~\ref{fig:image}; for Systems 1a and 2, which are barely detected in the \textit{HST} SNAPshot of eMACS\,J0252, we mark the image positions by overlaying contours of the Ly$\alpha$ flux from our MUSE observations.

%
% Table 1
%
\begin{table}
    \centering
\begin{supertabular}{lccl}
Image ID & R.A.\ & Dec.\ & $z_{\rm spec}$ \\
	& (J2000) & (J2000)  & \\[1mm]
\hline \\[-2mm]
 1.1 & 02 52 26.46 & $-$21 00 46.3 & \hspace{-1em}\rdelim\}{6}{*}[\;3.458]  \\
 1.2 & 02 52 26.55 & $-$21 00 50.4 &  \\
 1.3 & 02 52 28.61 & $-$21 01 12.8 &  \\
 1a.1 & 02 52 26.51 & $-$21 00 44.2 & \\
 1a.2 & 02 52 26.63 & $-$21 00 51.4 & \\
 1a.3 & 02 52 28.62 & $-$21 01 12.4 & \\
 2.1 & 02 52 26.93 & $-$21 01 05.7 & \hspace{-1em}\rdelim\}{3}{*}[\;4.148]  \\
 2.2 & 02 52 27.51 & $-$21 01 06.9 &  \\
 2.3 & 02 52 25.41 & $-$21 00 36.6 &  \\[1mm]
\hline
\end{supertabular}
\caption{Multiple-image families identified in the strong-lensing regime of eMACS\,J0252; see also Figs.~\ref{fig:image} (insets) and \ref{fig:SLspec}.}
    \label{tab:SL}
\end{table}

%
% Table 2
%
\begin{table*}
    \centering
    \begin{tabular}{cccccccc}
Potential & $\Delta\alpha$ & $\Delta\delta$ & $e$ & $\theta$ & r$_{\rm core}$ & r$_{\rm cut}$ & $\sigma$ \\
   & [arcsec] & [arcsec] & & [deg] & [kpc] & [kpc] & [km s$^{-1}$] \\[1mm]
\hline \\[-2mm]
Cluster & $  3.0^{+0.5}_{-0.3}$ & $ -3.8^{+0.3}_{-0.6}$ & $ 0.79^{+0.02}_{-0.02}$ & $130.4^{+0.3}_{-1.4}$ & $79^{+15}_{-4}$ & $[1000]$ & $1050^{+20}_{-18}$ \\[1mm]
BCG & $[0.0]$ & $[0.0]$ & $ 0.36^{+ 0.06}_{-0.37}$ & $165^{+ 11}_{-14}$ & $[0]$ & $126^{+153}_{-8}$ & $360^{+122}_{-19}$ \\[1mm]
$L^{\ast}$ galaxy & ... & ... & ... & ... & $[0.15]$ & $[45]$ & $[158]$\\[1mm]
\hline
   \end{tabular}
    \caption{Best-fit model parameters for the mass distribution of eMACS\,J0252. From left to right: mass component, position relative to the BCG ($\Delta\alpha$ and $\Delta\delta$), dPIE shape (ellipticity and orientation, $e$ and $\theta$), core and cut radii ($r_{\rm core}$ and $r_{\rm cut}$), and velocity dispersion ($\sigma$). The final row lists the global parameters adopted for a galaxy at the characteristic luminosity $L^\ast$, which is scaled to match the observed luminosity each of cluster member. Values in square brackets denote {\it a priori} settings that are kept fixed during the optimization setting. }
    \label{tab:massmodel}
\end{table*}

We use \textsc{Lenstool} \citep{2007NJPh....9..447J} to create a lens model of eMACS\,J0252. Adopting double Pseudo Isothermal Elliptical (dPIE) profiles for the mass distributions of both cluster- and galaxy-scale components, we optimise the shape and mass scaling parameters of the cluster-scale component and the central BCG to reproduce the three multiple-image systems. We include all spectroscopically confirmed cluster members as galaxy-scale halos that act as local perturbers of the large-scale gravitational potential. The best-fit model, characterized by the parameters listed in Table~\ref{tab:massmodel}, yields an rms offset of 0.09\arcsec\ between the predicted and observed locations of all images. Like the distribution of optical light from the BCG, the mass distribution of eMACS\,J0252 is characterized by an exceptionally high ellipticity; our lens model requires an axis ratio of $b/a=0.34\pm 0.01$ or
an ellipticity of $e=0.79$,\footnote{We define ellipticity as $e=\frac{1-(b/a)^2}{1+(b/a)^2}$, with $b/a$ being the axis ratio.}  a value far more extreme of the average of 0.52 found, e.g., from statistical gravitational-lensing analyses of SDSS clusters \citep[][]{2018MNRAS.475.2421S}.

%
% Figure 4
%
\begin{figure}
    \centering
    \includegraphics[width=0.5\textwidth]{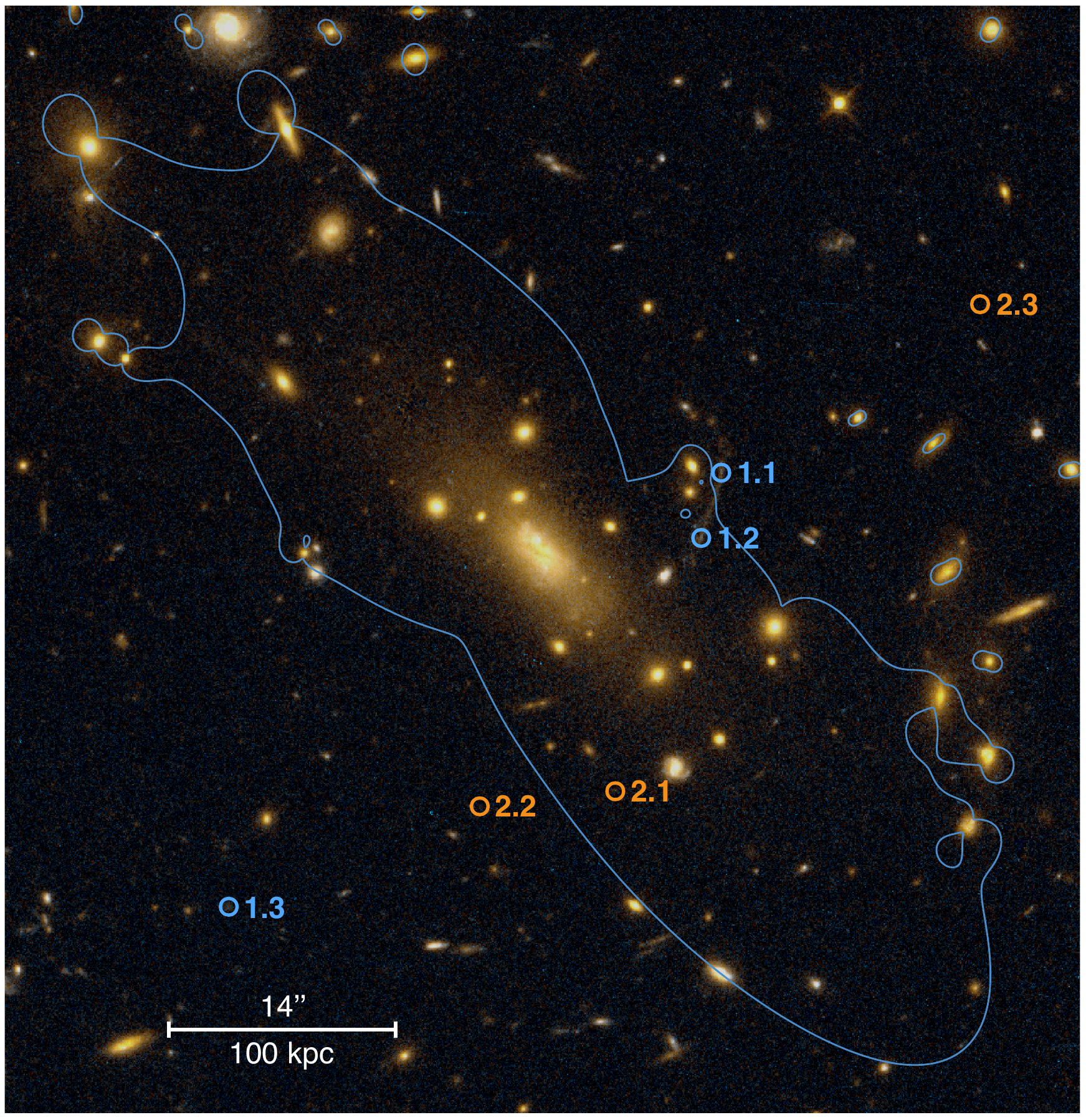}
    \caption{Critical line (for a source at $z=3.458$) of the lens model derived from the strong-lensing constraints listed in Table~\ref{tab:SL}, overlaid on the \textit{HST} image of eMACS\,J0252. For clarity we do not show the (very similar) critical line for the source redshift of multiple-image system 2. }
    \label{fig:SLmodel}
\end{figure}

The critical lines (for a source at $z=3.458$) from the best-fitting lens model are shown in  Fig.~\ref{fig:SLmodel}. The projected mass within a radius of 250\,kpc is $(1.94\pm0.02\pm0.1)\times 10^{14}$\,M$_\odot$ (statistical and estimated systematic uncertainties are listed separately); the total mass (not well constrained by strong-lensing measurements) exceeds $5\times 10^{14}$\,M$_\odot$. With an effective Einstein radius\footnote{We define the effective Einstein radius as $\sqrt{A_{\rm crit}/\pi}$, where $A_{\rm crit}$ is the area enclosed by the critical line.} of 19.9\arcsec\ for System 1 at $z=3.458$ (15.5\arcsec\ for a source at $z=2$), eMACS\,J0252 is a moderately strong gravitational lens, in agreement with the general trend for more powerful cluster lenses to be disturbed active mergers, extreme cases being MACSJ0717.5+3745 \citep[e.g.,][]{2016A&A...588A..99L} or Abell 2744 \citep[e.g.,][]{2015MNRAS.452.1437J}.

%
% Figure 5
%
\begin{figure}
\hspace*{-5mm}\includegraphics[width=0.52\textwidth]{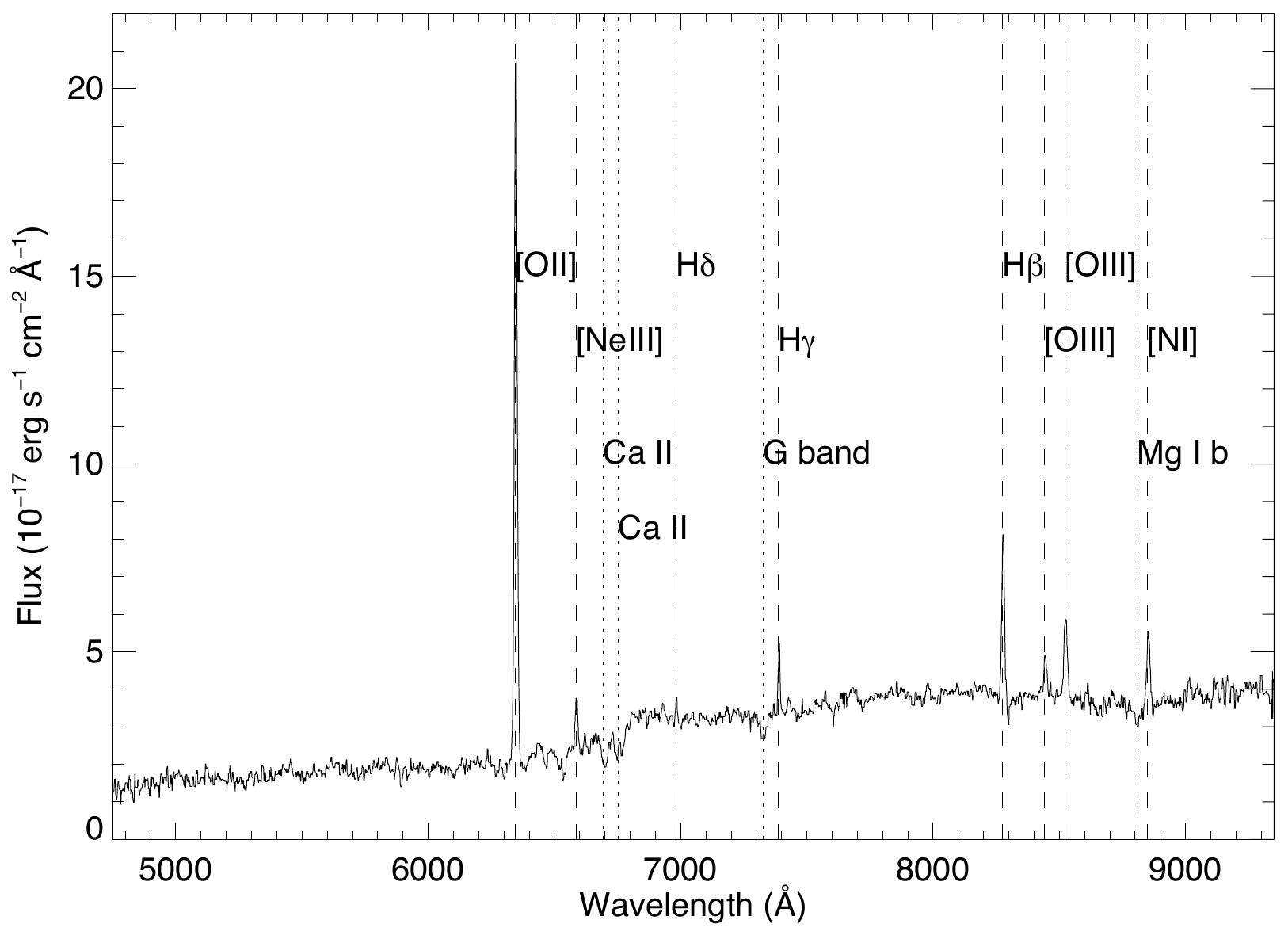}
\caption{Spectrum of the BCG of eMACS\,J0252, as recorded by MUSE and corrected for  Galactic extinction using $E(B-V)=0.028$ as measured by \citet{2011ApJ...737..103S}; notable emission and absorption features are labeled and marked by dashed and dotted lines, respectively. The extraction region of this spectrum is outlined in Fig.~\ref{fig:o2map}. 
\label{fig:bcgspec}
}
\end{figure}

\subsection{BCG properties}

\subsubsection{Optical photometry and morphology}

The cD galaxy ($\alpha_{\rm J2000}=$\,02$^{\rm h}$ 52$^{\rm m}$ 27.25$^{\rm s}$, $\delta_{\rm J2000}=-$21$^\circ$ 00$^\prime$ 51.2\arcsec; Figure \ref{fig:image}) in the core of eMACS\,J0252 is characterized by a prodigious luminosity ($M_{\rm g}=-24.1$), intersecting dust lanes, and an overall morphology that shows a strong resemblance to BCGs in cool-core clusters at lower redshift \citep{2015ApJ...805..177D}. The archival photometry from PanSTARRS-1, {\it GALEX} and {\it WISE} yields a luminosity of $5 \times 10^{11}$\,L$_\odot$ for the BCG in the rest-frame $i$-band and near-UV, optically blue colours similar to those of line-emitting BCGs at lower redshift \citep{2016MNRAS.461..560G}. 

Even more remarkable is the BCG's extreme elongation (see Figure \ref{fig:image} and \ref{fig:SLmodel}). Its ellipticity of $e=0.78$ (derived from an axis ratio of 0.35 measured from \textit{HST} imaging) is unrivaled among similarly massive clusters \citep[][]{2019MNRAS.490.4889H} and a $> 3 \sigma$ outlier compared to the average axis ratio of $0.7\pm0.1$ established for BCGs in general \citep{2011ApJS..195...15D}. % ($b/a=0.42$), 
Although the light envelopes of BCGs are known to become more elliptical with increasing radius, axis ratios below 0.4 are extremely rare and only observed at radii of at least 100 kpc \citep[][]{2020ApJS..247...43K}, far larger than the approximately 40 kpc probed by our shallow \textit{HST} data (Fig.~\ref{fig:o2map}).

\subsubsection{Radio properties}
\label{sec:radio}

The most notable feature of the bright compact radio source associated with the BCG is a very steep lower-frequency component ($\alpha\sim -2.0$) that dominates below 300\,MHz, and whose spectral slope is comparable to that of the massive lobed source in MS\,0735+74 \citep[$z=0.21$;][]{2005ApJ...620L...5C}; it also shares with MS\,0735+75  a similar rest-frame radio power of $1.3 \times 10^{25}$\,W\,Hz$^{-1}$. The radio emission above 1\,GHz has a significantly flatter spectrum ($\alpha \sim -0.5$), comparable to that of the compact-core emission found in all line-emitting BCGs \citep{2015MNRAS.453.1201H} (but notably missing in MS\,0735+74). In terms of radio power ($7.3 \times 10^{24}$\,W\,Hz$^{-1}$  at 10\,GHz), the core emission of the BCG in eMACS\,J0252 falls into the top 10 percentile of the distribution compiled by \citet{2015MNRAS.453.1201H} while the more extended, steep-spectrum emission still ranks in the top third. These properties suggest that accretion onto the active nucleus on the BCG is ongoing at present, after a sustained period of activity in the past that powered the aged, steep-spectrum lobes. A low-frequency observation at higher resolution of this source would help constrain the timing and energetics of this past outburst.

\subsubsection{Stellar populations and star formation}

%
% Figure 6
%
\begin{figure*}
    %\centering
    \includegraphics[width=0.23\textwidth]{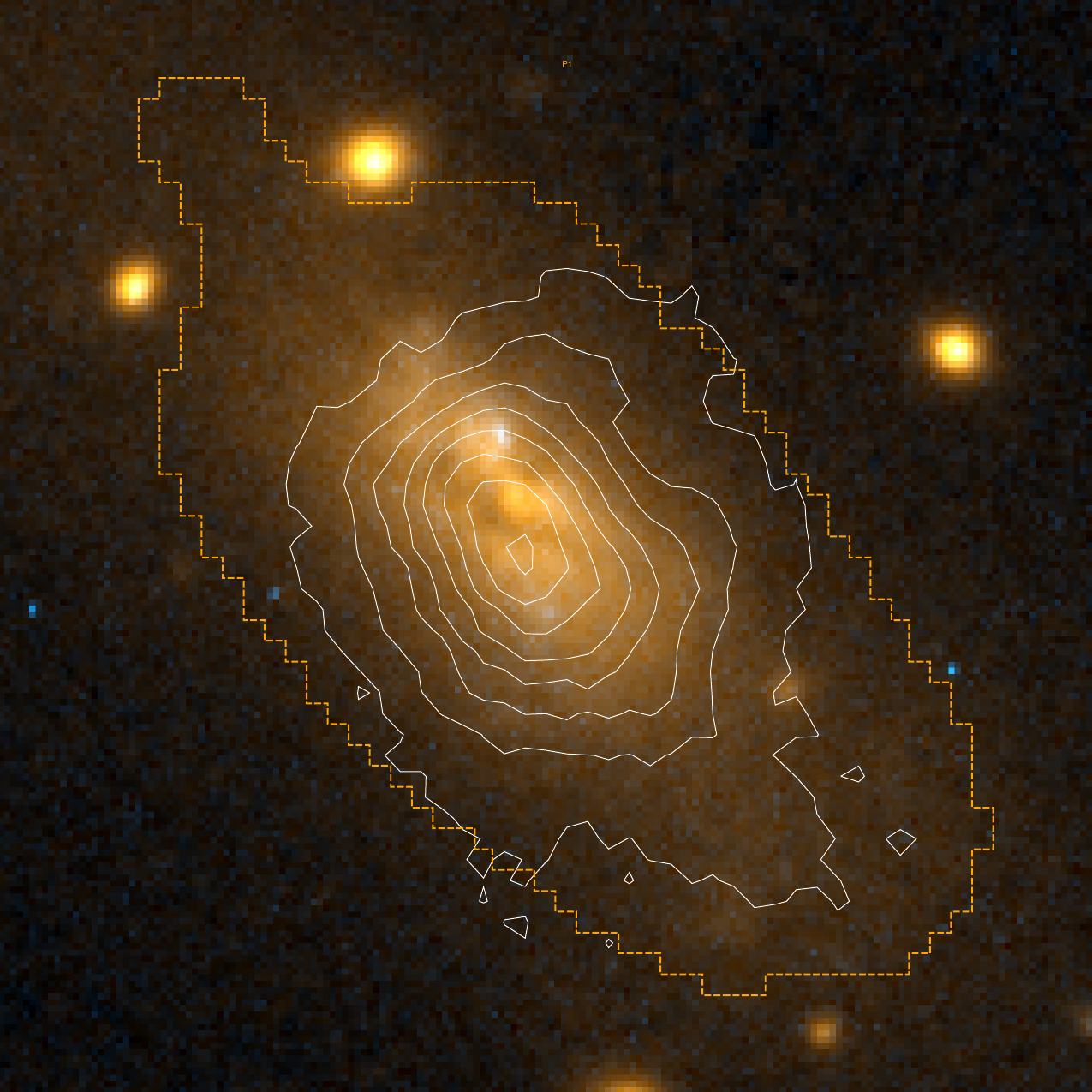}
    \includegraphics[width=0.23\textwidth]{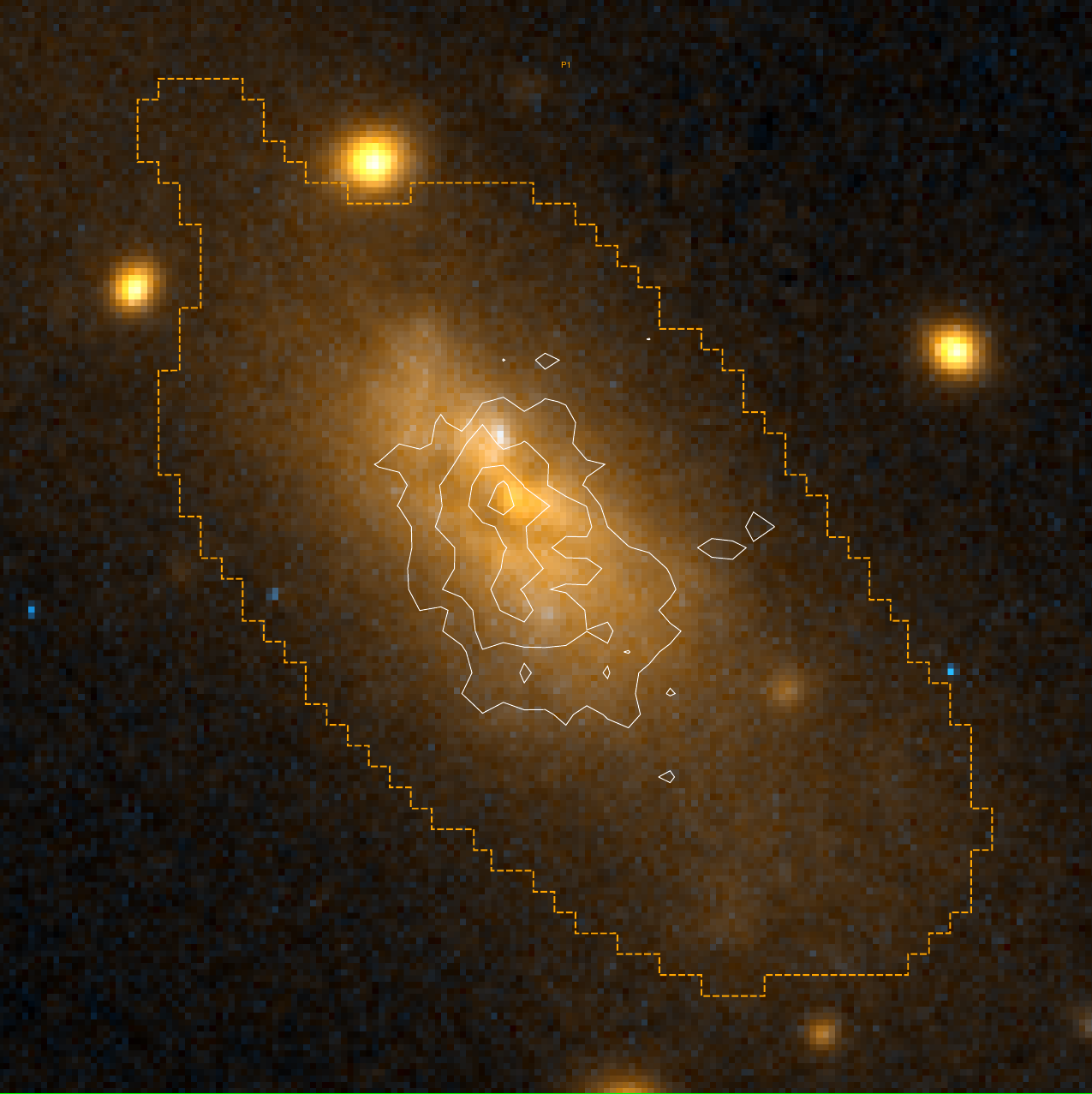}
    \includegraphics[width=0.23\textwidth]{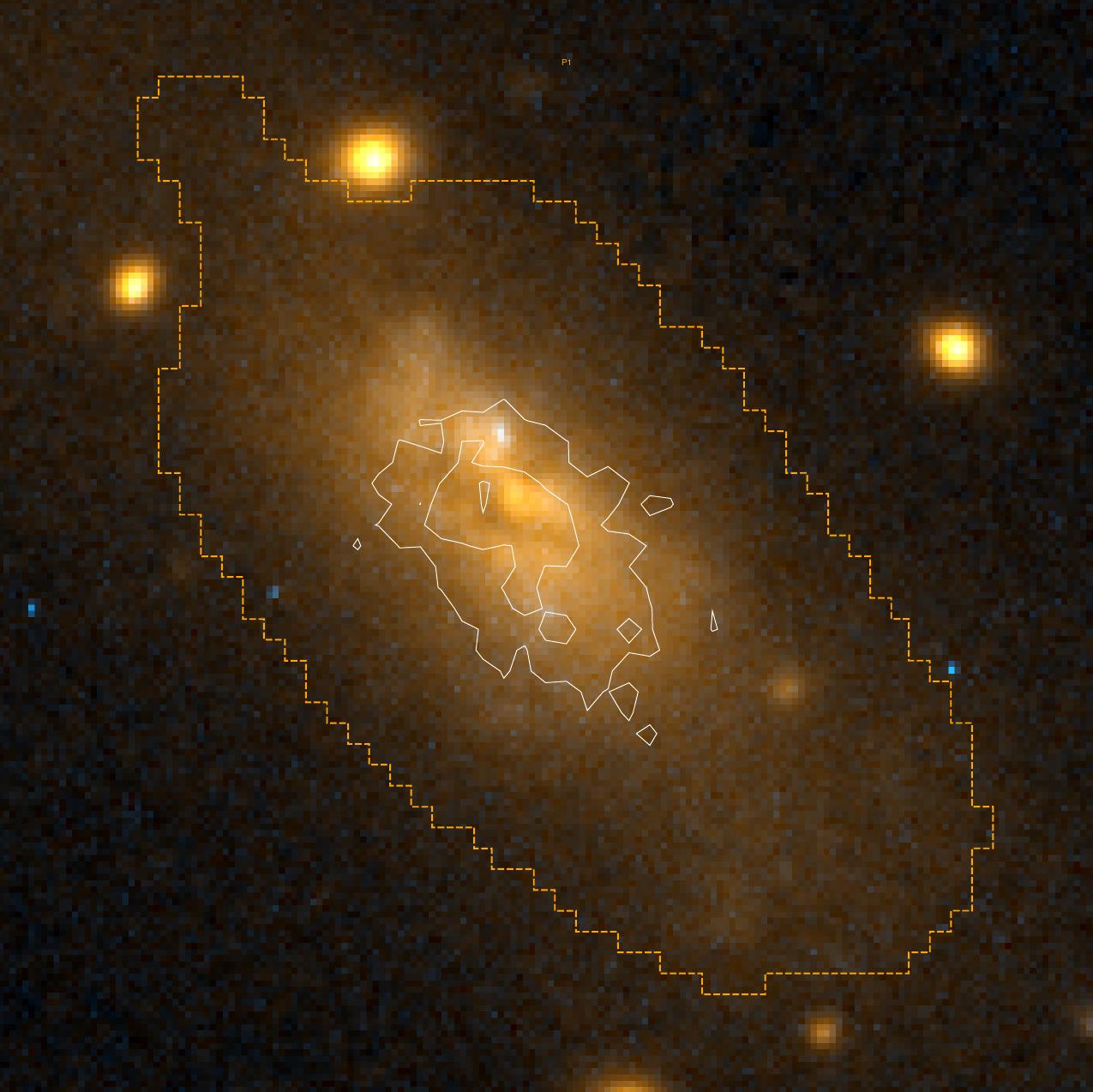}
     \includegraphics[width=0.23\textwidth]{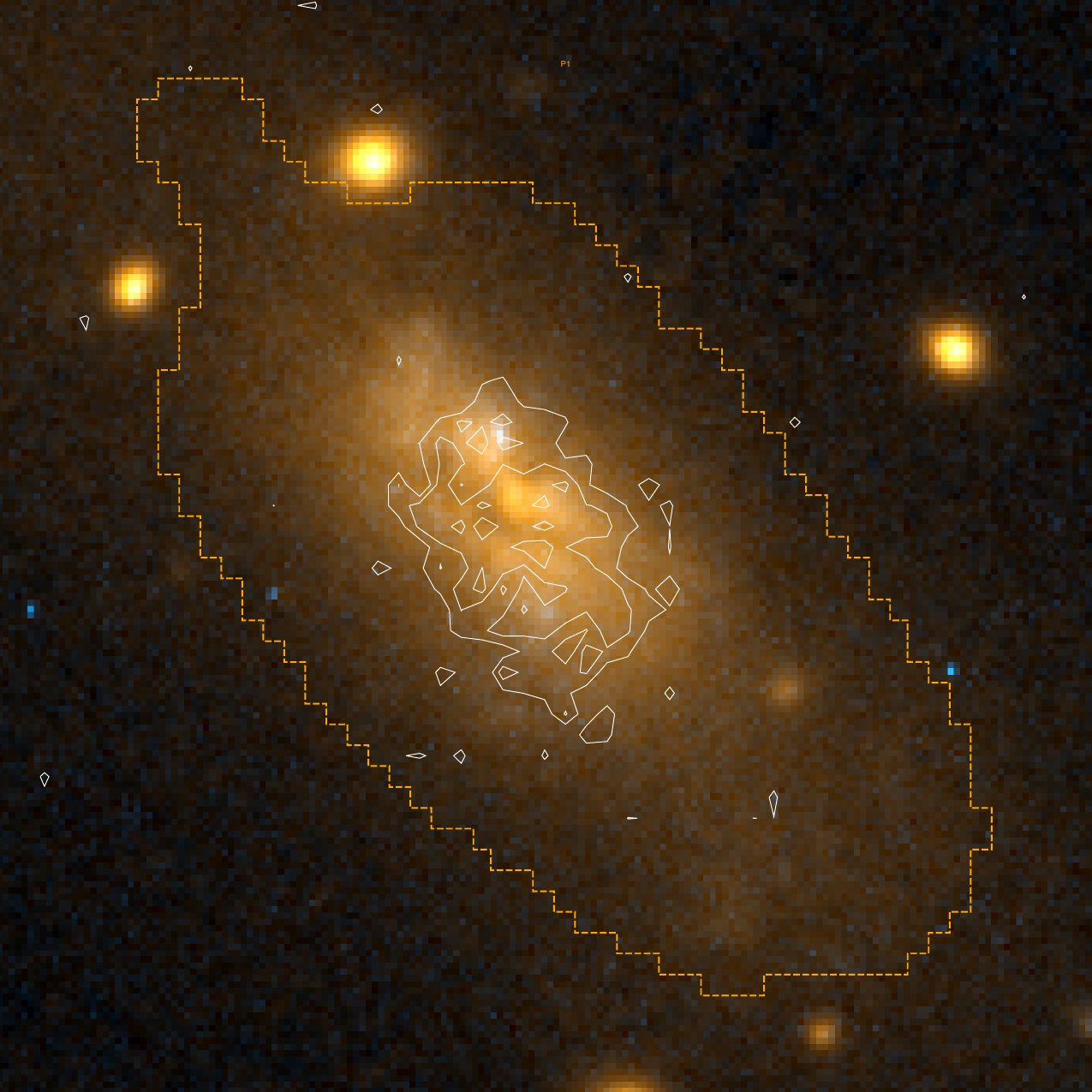}
    \caption{Contours (linearly spaced) of the emission-line flux from the BCG of eMACS\,J0252 as observed with MUSE, overlaid on the {\it HST} image shown in Fig.~\ref{fig:image}; the shown region spans 10.5\arcsec\ on a side (75 kpc at the cluster redshift). Left to right: [\ion{O}{II}], H$\beta$, [\ion{O}{III}], and [\ion{Ne}{III}].
    The region from which the BCG spectrum shown in Fig.~\ref{fig:bcgspec} was extracted is delineated by the dashed orange line.  }
    \label{fig:o2map}
\end{figure*}

Strong absorption features, such as \ion{Ca}{II} (H+K), the G band, and \ion{Mg}{I} b, in the spectrum of the BCG, as well as a pronounced Balmer break in the continuum, unequivocally establish the presence of an old stellar population, as found in all elliptical galaxies (Fig.~\ref{fig:bcgspec}). Rare in average ellipticals \citep[but not uncommon in BCGs;][and references therein]{2016MNRAS.461..560G} is the pronounced line emission seen in Fig.~\ref{fig:bcgspec}. 

\begin{figure}
\hspace*{-1mm}\includegraphics[width=0.48\textwidth]{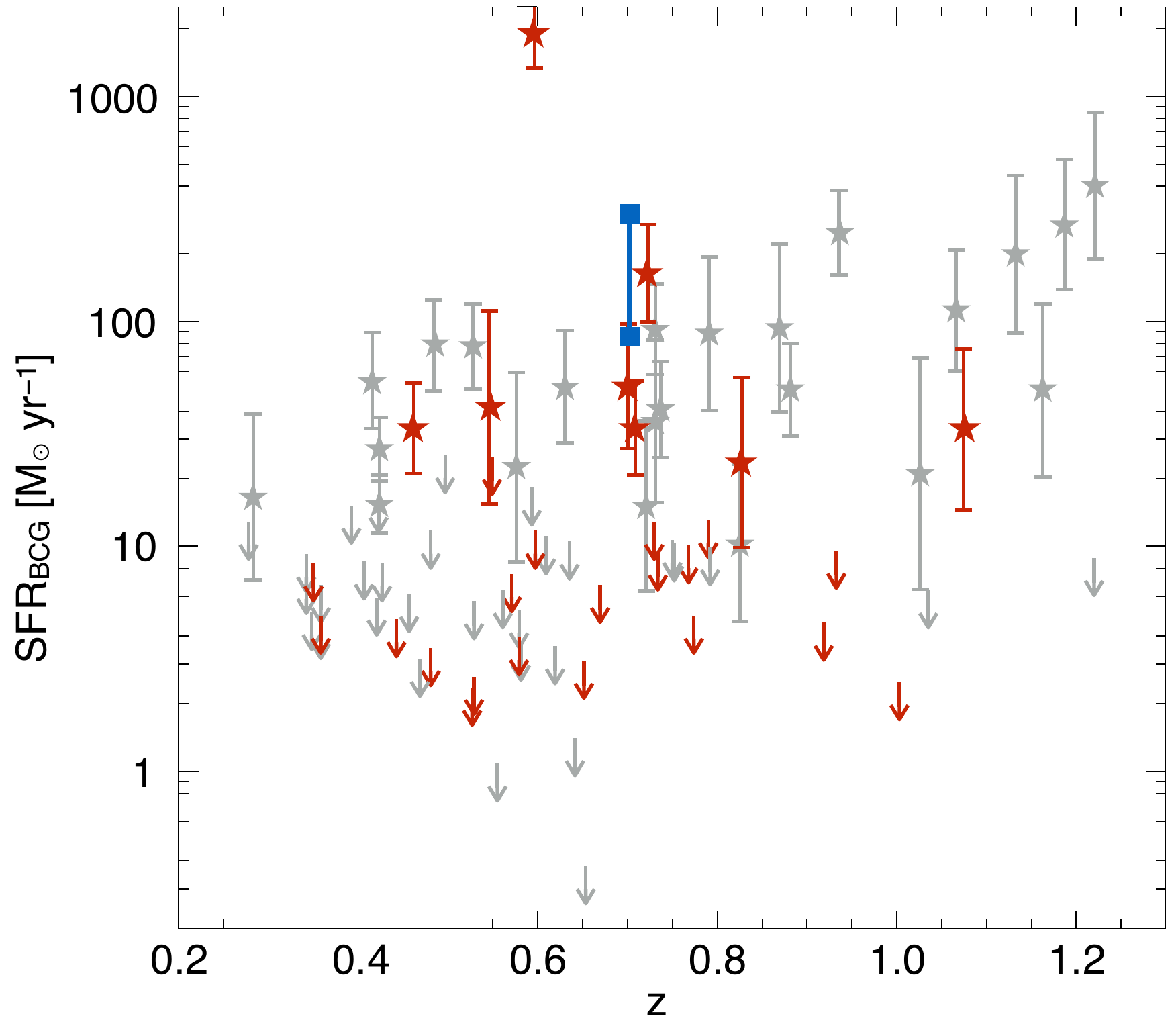}
\caption{Locus of the BCG of eMACS\,J0252 in the SFR vs redshift plane for BCGs \citep[adapted from][]{2016ApJ...817...86M}. Grey and red symbols represent BCGs in disturbed and relaxed clusters, respectively. Our data point, added in blue, is shown as a range from the lower to the upper limit, set by our MUSE and SCUBA2 observations, respectively.  \label{fig:sfr}}
\end{figure}

From the [\ion{O}{II}]$\lambda 3727$\AA\ emission-line flux recorded in our MUSE observation we measure a high star-formation rate of 85 M$_\odot$ yr$^{-1}$, following \citet{1989ApJ...344..685K}. Although this value accounts for both Galactic extinction and intrinsic extinction as estimated from the observed Balmer decrement, the latter is flux-weighted and thus biased against high-extinction regions. Indeed the corresponding $E(B-V)$ value of about 0.05 appears unphysically low for the derived high intrinsic [\ion{O}{II}]$\lambda 3727$\AA\ emission-line luminosity of $6.1\times 10^{42}$ erg s$^{-1}$, a combination that is in stark conflict with the correlation found by \citet{2004AJ....127.2002K}. Considering, in addition, the compelling evidence of a dust lane (apparent in the short \textit{HST} SNAPshot shown in Fig.~\ref{fig:image}), we therefore consider the quoted intrinsic extinction and the corresponding star-formation rate lower limits. The lack of a detection in the SCUBA2 observations, however, places a corresponding 3-$\sigma$ upper limit of  300\,M$_\odot$\,yr$^{-1}$ on the BCG  star-formation rate. Both limits are shown in Fig.~\ref{fig:sfr} and identify the BCG of eMACS\,J0252 as an extreme system in terms of star-formation rate. 

We note that a contribution to the observed [\ion{O}{II}] flux from nuclear activity can not be ruled out. Not only do the system's radio properties suggest the presence of an AGN, the ratio of the fluxes of the [\ion{O}{III}] and H$\beta$ emission lines of $\log([\ion{O}{III}]/{\rm H}\beta) = -0.25$ also places the BCG of eMACS\,J0252 in the composite region of the BPT diagram \citep{1981PASP...93....5B}. A correction for any AGN contribution to the [\ion{O}{II}] flux (and hence the deduced star-formation rate) will require near-infrared spectroscopy to determine the H$\alpha$ emission-line flux.

\subsubsection{Galaxy dynamics}

Using the pPXF code of \citet{2004PASP..116..138C}, we measure the central stellar velocity dispersion of the BCG to be ($380\pm 30$) km s$^{-1}$, well above the average value of 260 km/s for BCGs in the local Universe \citep{2014ApJ...797...82L} and high for BCGs at any redshift \citep{2020ApJ...891..129S}.

\begin{figure}
    \centering
    \includegraphics[width=0.5\textwidth]{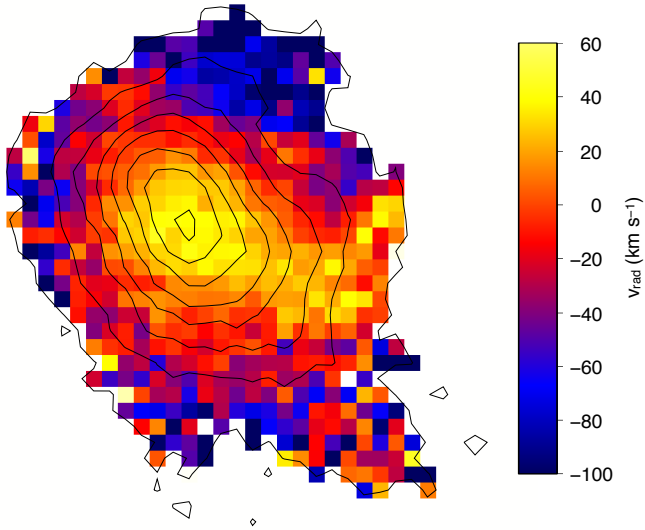}
    \caption{Radial velocity distribution derived from the observed wavelength of the [\ion{O}{II}] line, as recorded by MUSE. The overlaid contours show the [\ion{O}{II}] line flux for reference and scale (see also Fig.~\ref{fig:o2map}).     }
    \label{fig:velmap}
\end{figure}

The strong [\ion{O}{II}] emission from the BCG (Fig.~\ref{fig:bcgspec}) extends to over 20 kpc in radius from the galactic centre, as shown in Fig.~\ref{fig:o2map}. In order to constrain the BCG's kinematics, we track the spectral location and shape of the [\ion{O}{II}] line across the full extent of this region. We interpret the recorded shifts of the centroid of the [\ion{O}{II}] line as being caused by changes in the radial component of the peculiar velocity of the ISM, noting that variations in the ratio of the strengths of the 3726.0\AA\ and 3728.7\AA\ lines that make up the [\ion{O}{II}] doublet\footnote{Such variations would be indicative of changes in the electron density \citep[][]{1974agn..book.....O}.} can not account for the observed spectral shifts. The resulting map of the BCG's radial-velocity field is shown in Fig.~\ref{fig:velmap}. Unlike some line-emitting BCGs at lower redshift \citep{2016MNRAS.460.1758H}, the BCG of eMACS\,J0252 shows no pronounced velocity asymmetry indicative of rotation or bulk motions associated with ``sloshing" \citep[first described by][]{2000ApJ...541..542M}, but instead a  radial gradient of $(90\pm 30)$ km s$^{-1}$ amplitude that is  consistent with radial infall of gas: the highest (positive) radial velocities are observed near the galactic centre, and no motion is detected at the largest radii (in projection) where infall is perpendicular to our line of sight. The timescale for infall from 30~kpc at this velocity is 330~Myr, in good qualitative agreement with the cooling times of about 1 Gyr observed for fully relaxed cool-core clusters \citep[e.g.,][]{2010A&A...513A..37H}.

The above interpretation of the radial-velocity pattern shown in Fig.~\ref{fig:velmap} as being indicative of isotropic infall requires the gas within the shown region to be optically thick\footnote{If the gas around the BCG were optically thin, like the ICM over most of the cluster volume, [\ion{O}{II}] emission would be detectable both from gas in front and behind the BCG; as a result, isotropic infall would create no net shift of the [\ion{O}{ii}] line but only an effective broadening.}. Although speculative for eMACS\,J0252 at this point, this assumption is consistent with the high reddening values of $E(B-V)\approx 0.6$ measured in the centres of massive cool-core clusters \citep[][]{2010ApJ...719.1619O,2013ApJ...765L..37M}  and excess absorption equivalent to neutral-hydrogen column densities of several $10^{21}$ cm$^{-2}$ \citep[e.g.,][]{2019ApJ...885...63M}.
%
% Figure 7
%

\section{Discussion and Conclusions}
\label{sec:summary}

As galaxy clusters grow through mergers and accretion, so do the giant ellipticals in their cores.
The physical processes involved are closely intertwined, and the timescales involved are huge: it takes Gyrs for a cluster to dynamically relax after a merger, and it takes Gyrs for extremely luminous and extended cD galaxies to form in their centers. As a result, discoveries of highly evolved systems at high redshift offer rare opportunities to constrain the timeline of structure formation and growth at large look-back times.

%
% Figure 8
%

The discovery of eMACS\,J0252 represents such an opportunity. At the system's redshift of $z=0.703$, the Universe is only a little over half its present age old, and yet the BCG of eMACS\,J0252 appears to have already evolved into one of the most massive galaxies known at \textit{any} epoch. Specifically, eMACS\,J0252 and its BCG meet all of the following criteria for highly evolved, massive systems:

\begin{itemize}
    \item Presence of a single, extremely luminous BCG;
    \item Excellent alignment between the major axes of the BCG and the overall galaxy distribution;
    \item High cluster velocity dispersion (over 1,000 km s$^{-1}$), high X-ray luminosity (over $1\times 10^{45}$ erg s$^{-1}$), and high central mass ($M(r<250\, \rm{kpc})\sim 2\times 10^{14}$ M$_\odot$);
    \item No evidence of substructure along the line of sight;
    \item A prodigiously star forming BCG that, at comparable redshifts, is only surpassed by the Phoenix Cluster ($z=0.597$);
    \item Evidence of a radial flow of gas within 10 to 20 kpc of the BCG, traced by variations in the observed wavelength of the [\ion{O}{II}] line;
    \item Presence of radio emission suggestive of sustained nuclear activity characteristic of AGN feedback in cool-core clusters.
\end{itemize}

Of the characteristics listed above, one in particular stands out: the detection of a radial flow of star-forming gas in the immediate vicinity of the BCG. Although the observed velocity pattern is fully consistent with isotropic infall, this interpretation holds only if the gas in the immediate vicinity ($r\leq 20$ kpc) is optically thick, a speculative assumption that is at present only supported by the potentially high density of the ICM in the very centre of cool cluster cores and the presence of visible dust lanes across the BCG of eMACS\,J0252. If confirmed as an isotropic flow by in-depth follow-up studies, the signature of radial infall seen in Fig.~\ref{fig:velmap} (expected from, and fully consistent with, extensive studies of feedback processes in the cores of relaxed clusters \citep[for a recent review see, e.g.,][]{2012NJPh...14e5023M}) would constitute, to the best of our knowledge, the first direct kinematic confirmation of the cooling-flow picture advanced by \citet{1984Natur.310..733F}.

In addition, the extreme ellipticity of both the BCG and the cluster-scale dark-matter distribution of its host cluster (as constrained by our strong-lensing mass model) deserves special emphasis. At $e=0.8$, the ellipticity of eMACSJ0252 provides a rare glimpse of the extreme end of the cluster-halo shape distribution for comparison with predictions from numerical simulations \citep{2020MNRAS.tmp.2971H}.

Although the existing observational data have established eMACS\,J0252 unambiguously as an exceptional cluster, dominated by an equally exceptional BCG, our knowledge of the system remains preliminary in several important areas. Specifically the measurement of key characteristics of the BCG (AGN properties, gas and dust content, morphology, size of core / nuclei, full extent and profile of diffuse halo, etc) is hampered by the limited depth and spectral coverage of the existing data. Even more important, however, will be a robust determination of the fundamental properties and dynamic state of the BCG's cluster environment characterized, for instance, by the presence of a prominent cool core, a task that will require dedicated follow-up observations across the electromagnetic spectrum, in particular high-resolution X-ray imaging spectroscopy of the intracluster medium.

\section*{Acknowledgements}

IRS and ACE acknowledge support from STFC (ST/T000244/1).
Based on observations obtained at the international Gemini Observatory, a program of NSF’s NOIRLab, which is managed by the Association of Universities for Research in Astronomy (AURA) under a cooperative agreement with the National Science Foundation on behalf of the Gemini Observatory partnership: the National Science Foundation (United States), National Research Council (Canada), Agencia Nacional de Investigaci\'{o}n y Desarrollo (Chile), Ministerio de Ciencia, Tecnolog\'{i}a e Innovaci\'{o}n (Argentina), Minist\'{e}rio da Ci\^{e}ncia, Tecnologia, Inova\c{c}\~{o}es e Comunica\c{c}\~{o}es (Brazil), and Korea Astronomy and Space Science Institute (Republic of Korea).
Also based on observations made with the NASA/ESA {\it Hubble Space Telescope}, obtained from the Data Archive at the Space Telescope Science Institute, which is operated by the Association of Universities for Research in Astronomy, Inc., under NASA contract NAS 5-26555. These observations are associated with program GO-15307. Some of the data presented herein were obtained at the W.M.\ Keck Observatory, which is operated as a scientific partnership among the California Institute of Technology, the University of California, and the National Aeronautics and Space Administration. The Observatory was made possible by the generous financial support of the W.M.\ Keck Foundation. The authors wish to recognize and acknowledge the very significant cultural role and reverence that the summit of Maunakea has always had within the indigenous Hawaiian community.  We are most fortunate to have the opportunity to conduct observations from this mountain.

\section*{Data Availability}
The data underlying this article are available in the MAST, VLT, and Keck data archives.

\bibliography{paper_bib.bib}

\begin{thebibliography}{}
\makeatletter
\relax
\def\mn@urlcharsother{\let\do\@makeother \do\$\do\&\do\#\do\^\do\_\do\%\do\~}
\def\mn@doi{\begingroup\mn@urlcharsother \@ifnextchar [ {\mn@doi@}
  {\mn@doi@[]}}
\def\mn@doi@[#1]#2{\def\@tempa{#1}\ifx\@tempa\@empty \href
  {http://dx.doi.org/#2} {doi:#2}\else \href {http://dx.doi.org/#2} {#1}\fi
  \endgroup}
\def\mn@eprint#1#2{\mn@eprint@#1:#2::\@nil}
\def\mn@eprint@arXiv#1{\href {http://arxiv.org/abs/#1} {{\tt arXiv:#1}}}
\def\mn@eprint@dblp#1{\href {http://dblp.uni-trier.de/rec/bibtex/#1.xml}
  {dblp:#1}}
\def\mn@eprint@#1:#2:#3:#4\@nil{\def\@tempa {#1}\def\@tempb {#2}\def\@tempc
  {#3}\ifx \@tempc \@empty \let \@tempc \@tempb \let \@tempb \@tempa \fi \ifx
  \@tempb \@empty \def\@tempb {arXiv}\fi \@ifundefined
  {mn@eprint@\@tempb}{\@tempb:\@tempc}{\expandafter \expandafter \csname
  mn@eprint@\@tempb\endcsname \expandafter{\@tempc}}}

\bibitem[\protect\citeauthoryear{{Albert}, {White}  \& {Morgan}}{{Albert}
  et~al.}{1977}]{1977ApJ...211..309A}
{Albert} C.~E.,  {White} R.~A.,   {Morgan} W.~W.,  1977, \mn@doi [\apj]
  {10.1086/154935}, \href
  {https://ui.adsabs.harvard.edu/abs/1977ApJ...211..309A} {211, 309}

\bibitem[\protect\citeauthoryear{{Baldwin}, {Phillips}  \&
  {Terlevich}}{{Baldwin} et~al.}{1981}]{1981PASP...93....5B}
{Baldwin} J.~A.,  {Phillips} M.~M.,   {Terlevich} R.,  1981, \mn@doi [\pasp]
  {10.1086/130766}, \href
  {https://ui.adsabs.harvard.edu/abs/1981PASP...93....5B} {93, 5}

\bibitem[\protect\citeauthoryear{{Beers} \& {Geller}}{{Beers} \&
  {Geller}}{1983}]{1983ApJ...274..491B}
{Beers} T.~C.,  {Geller} M.~J.,  1983, \mn@doi [\apj] {10.1086/161463}, \href
  {https://ui.adsabs.harvard.edu/abs/1983ApJ...274..491B} {274, 491}

\bibitem[\protect\citeauthoryear{{Bellstedt} et~al.,}{{Bellstedt}
  et~al.}{2016}]{2016MNRAS.460.2862B}
{Bellstedt} S.,  et~al., 2016, \mn@doi [\mnras] {10.1093/mnras/stw1184}, \href
  {https://ui.adsabs.harvard.edu/abs/2016MNRAS.460.2862B} {460, 2862}

\bibitem[\protect\citeauthoryear{{Bleem} et~al.,}{{Bleem}
  et~al.}{2020}]{2020ApJS..247...25B}
{Bleem} L.~E.,  et~al., 2020, \mn@doi [\apjs] {10.3847/1538-4365/ab6993}, \href
  {https://ui.adsabs.harvard.edu/abs/2020ApJS..247...25B} {247, 25}

\bibitem[\protect\citeauthoryear{{Bonaventura} et~al.,}{{Bonaventura}
  et~al.}{2017}]{2017MNRAS.469.1259B}
{Bonaventura} N.~R.,  et~al., 2017, \mn@doi [\mnras] {10.1093/mnras/stx722},
  \href {https://ui.adsabs.harvard.edu/abs/2017MNRAS.469.1259B} {469, 1259}

\bibitem[\protect\citeauthoryear{{Cappellari} \& {Emsellem}}{{Cappellari} \&
  {Emsellem}}{2004}]{2004PASP..116..138C}
{Cappellari} M.,  {Emsellem} E.,  2004, \mn@doi [\pasp] {10.1086/381875}, \href
  {https://ui.adsabs.harvard.edu/abs/2004PASP..116..138C} {116, 138}

\bibitem[\protect\citeauthoryear{{Cohen}, {Clarke}, {Feretti}  \&
  {Kassim}}{{Cohen} et~al.}{2005}]{2005ApJ...620L...5C}
{Cohen} A.~S.,  {Clarke} T.~E.,  {Feretti} L.,   {Kassim} N.~E.,  2005, \mn@doi
  [\apjl] {10.1086/428572}, \href
  {https://ui.adsabs.harvard.edu/abs/2005ApJ...620L...5C} {620, L5}

\bibitem[\protect\citeauthoryear{{Coziol}, {Andernach}, {Caretta},
  {Alamo-Mart{\'\i}nez}  \& {Tago}}{{Coziol}
  et~al.}{2009}]{2009AJ....137.4795C}
{Coziol} R.,  {Andernach} H.,  {Caretta} C.~A.,  {Alamo-Mart{\'\i}nez} K.~A.,
  {Tago} E.,  2009, \mn@doi [\aj] {10.1088/0004-6256/137/6/4795}, \href
  {https://ui.adsabs.harvard.edu/abs/2009AJ....137.4795C} {137, 4795}

\bibitem[\protect\citeauthoryear{{De Lucia} \& {Blaizot}}{{De Lucia} \&
  {Blaizot}}{2007}]{2007MNRAS.375....2D}
{De Lucia} G.,  {Blaizot} J.,  2007, \mn@doi [\mnras]
  {10.1111/j.1365-2966.2006.11287.x}, \href
  {https://ui.adsabs.harvard.edu/abs/2007MNRAS.375....2D} {375, 2}

\bibitem[\protect\citeauthoryear{{Donahue} et~al.,}{{Donahue}
  et~al.}{2015}]{2015ApJ...805..177D}
{Donahue} M.,  et~al., 2015, \mn@doi [\apj] {10.1088/0004-637X/805/2/177},
  \href {https://ui.adsabs.harvard.edu/abs/2015ApJ...805..177D} {805, 177}

\bibitem[\protect\citeauthoryear{{Donzelli}, {Muriel}  \& {Madrid}}{{Donzelli}
  et~al.}{2011}]{2011ApJS..195...15D}
{Donzelli} C.~J.,  {Muriel} H.,   {Madrid} J.~P.,  2011, \mn@doi [\apjs]
  {10.1088/0067-0049/195/2/15}, \href
  {https://ui.adsabs.harvard.edu/abs/2011ApJS..195...15D} {195, 15}

\bibitem[\protect\citeauthoryear{{Dressler}}{{Dressler}}{1984}]{1984ARA&A..22..185D}
{Dressler} A.,  1984, \mn@doi [\araa] {10.1146/annurev.astro.22.1.185}, \href
  {https://ui.adsabs.harvard.edu/abs/1984ARA&A..22..185D} {22, 185}

\bibitem[\protect\citeauthoryear{{Ebeling} et~al.,}{{Ebeling}
  et~al.}{2013}]{2013MNRAS.432...62E}
{Ebeling} H.,  et~al., 2013, \mn@doi [\mnras] {10.1093/mnras/stt387}, \href
  {https://ui.adsabs.harvard.edu/abs/2013MNRAS.432...62E} {432, 62}

\bibitem[\protect\citeauthoryear{{Faber} et~al.,}{{Faber}
  et~al.}{2003}]{2003SPIE.4841.1657F}
{Faber} S.~M.,  et~al., 2003, in {Iye} M.,  {Moorwood} A. F.~M.,  eds,  Society
  of Photo-Optical Instrumentation Engineers (SPIE) Conference Series Vol.
  4841, Instrument Design and Performance for Optical/Infrared Ground-based
  Telescopes. pp 1657--1669, \mn@doi{10.1117/12.460346}

\bibitem[\protect\citeauthoryear{{Fabian}, {Nulsen}  \& {Canizares}}{{Fabian}
  et~al.}{1984}]{1984Natur.310..733F}
{Fabian} A.~C.,  {Nulsen} P.~E.~J.,   {Canizares} C.~R.,  1984, \mn@doi [\nat]
  {10.1038/310733a0}, \href
  {https://ui.adsabs.harvard.edu/abs/1984Natur.310..733F} {310, 733}

\bibitem[\protect\citeauthoryear{{Gladders} et~al.,}{{Gladders}
  et~al.}{2019}]{2019hst..prop16017G}
{Gladders} M.~D.,  et~al., 2019, {Building the SPT-HST Legacy: Imaging Massive
  Clusters to z=1.5}, HST Proposal

\bibitem[\protect\citeauthoryear{{Green} et~al.,}{{Green}
  et~al.}{2016}]{2016MNRAS.461..560G}
{Green} T.~S.,  et~al., 2016, \mn@doi [\mnras] {10.1093/mnras/stw1338}, \href
  {https://ui.adsabs.harvard.edu/abs/2016MNRAS.461..560G} {461, 560}

\bibitem[\protect\citeauthoryear{{Hamer} et~al.,}{{Hamer}
  et~al.}{2016}]{2016MNRAS.460.1758H}
{Hamer} S.~L.,  et~al., 2016, \mn@doi [\mnras] {10.1093/mnras/stw1054}, \href
  {https://ui.adsabs.harvard.edu/abs/2016MNRAS.460.1758H} {460, 1758}

\bibitem[\protect\citeauthoryear{{Harvey}, {Robertson}, {Tam}, {Jauzac},
  {Massey}, {Rhodes}  \& {McCarthy}}{{Harvey}
  et~al.}{2020}]{2020MNRAS.tmp.2971H}
{Harvey} D.,  {Robertson} A.,  {Tam} S.-I.,  {Jauzac} M.,  {Massey} R.,
  {Rhodes} J.,   {McCarthy} I.~G.,  2020, \mn@doi [\mnras]
  {10.1093/mnras/staa3193}, \href
  {https://ui.adsabs.harvard.edu/abs/2020MNRAS.tmp.2971H} {}

\bibitem[\protect\citeauthoryear{{Herbonnet}, {von der Linden}, {Allen},
  {Mantz}, {Modumudi}, {Morris}  \& {Kelly}}{{Herbonnet}
  et~al.}{2019}]{2019MNRAS.490.4889H}
{Herbonnet} R.,  {von der Linden} A.,  {Allen} S.~W.,  {Mantz} A.~B.,
  {Modumudi} P.,  {Morris} R.~G.,   {Kelly} P.~L.,  2019, \mn@doi [\mnras]
  {10.1093/mnras/stz2913}, \href
  {https://ui.adsabs.harvard.edu/abs/2019MNRAS.490.4889H} {490, 4889}

\bibitem[\protect\citeauthoryear{{Hogan} et~al.,}{{Hogan}
  et~al.}{2015}]{2015MNRAS.453.1201H}
{Hogan} M.~T.,  et~al., 2015, \mn@doi [\mnras] {10.1093/mnras/stv1517}, \href
  {https://ui.adsabs.harvard.edu/abs/2015MNRAS.453.1201H} {453, 1201}

\bibitem[\protect\citeauthoryear{{Hudson}, {Mittal}, {Reiprich}, {Nulsen},
  {Andernach}  \& {Sarazin}}{{Hudson} et~al.}{2010}]{2010A&A...513A..37H}
{Hudson} D.~S.,  {Mittal} R.,  {Reiprich} T.~H.,  {Nulsen} P.~E.~J.,
  {Andernach} H.,   {Sarazin} C.~L.,  2010, \mn@doi [\aap]
  {10.1051/0004-6361/200912377}, \href
  {https://ui.adsabs.harvard.edu/abs/2010A&A...513A..37H} {513, A37}

\bibitem[\protect\citeauthoryear{{Jauzac} et~al.,}{{Jauzac}
  et~al.}{2015}]{2015MNRAS.452.1437J}
{Jauzac} M.,  et~al., 2015, \mn@doi [\mnras] {10.1093/mnras/stv1402}, \href
  {https://ui.adsabs.harvard.edu/abs/2015MNRAS.452.1437J} {452, 1437}

\bibitem[\protect\citeauthoryear{{Jullo}, {Kneib}, {Limousin},
  {El{\'\i}asd{\'o}ttir}, {Marshall}  \& {Verdugo}}{{Jullo}
  et~al.}{2007}]{2007NJPh....9..447J}
{Jullo} E.,  {Kneib} J.~P.,  {Limousin} M.,  {El{\'\i}asd{\'o}ttir} {\'A}.,
  {Marshall} P.~J.,   {Verdugo} T.,  2007, \mn@doi [New Journal of Physics]
  {10.1088/1367-2630/9/12/447}, \href
  {https://ui.adsabs.harvard.edu/abs/2007NJPh....9..447J} {9, 447}

\bibitem[\protect\citeauthoryear{{Kennicutt}}{{Kennicutt}}{1989}]{1989ApJ...344..685K}
{Kennicutt} Robert~C. J.,  1989, \mn@doi [\apj] {10.1086/167834}, \href
  {https://ui.adsabs.harvard.edu/abs/1989ApJ...344..685K} {344, 685}

\bibitem[\protect\citeauthoryear{{Kewley}, {Geller}  \& {Jansen}}{{Kewley}
  et~al.}{2004}]{2004AJ....127.2002K}
{Kewley} L.~J.,  {Geller} M.~J.,   {Jansen} R.~A.,  2004, \mn@doi [\aj]
  {10.1086/382723}, \href
  {https://ui.adsabs.harvard.edu/abs/2004AJ....127.2002K} {127, 2002}

\bibitem[\protect\citeauthoryear{{Kluge} et~al.,}{{Kluge}
  et~al.}{2020}]{2020ApJS..247...43K}
{Kluge} M.,  et~al., 2020, \mn@doi [\apjs] {10.3847/1538-4365/ab733b}, \href
  {https://ui.adsabs.harvard.edu/abs/2020ApJS..247...43K} {247, 43}

\bibitem[\protect\citeauthoryear{{Koekemoer} et~al.,}{{Koekemoer}
  et~al.}{2011}]{2011ApJS..197...36K}
{Koekemoer} A.~M.,  et~al., 2011, \mn@doi [\apjs] {10.1088/0067-0049/197/2/36},
  \href {https://ui.adsabs.harvard.edu/abs/2011ApJS..197...36K} {197, 36}

\bibitem[\protect\citeauthoryear{{Kormendy} \& {Djorgovski}}{{Kormendy} \&
  {Djorgovski}}{1989}]{1989ARA&A..27..235K}
{Kormendy} J.,  {Djorgovski} S.,  1989, \mn@doi [\araa]
  {10.1146/annurev.aa.27.090189.001315}, \href
  {https://ui.adsabs.harvard.edu/abs/1989ARA&A..27..235K} {27, 235}

\bibitem[\protect\citeauthoryear{{Lauer}, {Postman}, {Strauss}, {Graves}  \&
  {Chisari}}{{Lauer} et~al.}{2014}]{2014ApJ...797...82L}
{Lauer} T.~R.,  {Postman} M.,  {Strauss} M.~A.,  {Graves} G.~J.,   {Chisari}
  N.~E.,  2014, \mn@doi [\apj] {10.1088/0004-637X/797/2/82}, \href
  {https://ui.adsabs.harvard.edu/abs/2014ApJ...797...82L} {797, 82}

\bibitem[\protect\citeauthoryear{{Limousin} et~al.,}{{Limousin}
  et~al.}{2016}]{2016A&A...588A..99L}
{Limousin} M.,  et~al., 2016, \mn@doi [\aap] {10.1051/0004-6361/201527638},
  \href {https://ui.adsabs.harvard.edu/abs/2016A&A...588A..99L} {588, A99}

\bibitem[\protect\citeauthoryear{{Lin} \& {Mohr}}{{Lin} \&
  {Mohr}}{2004}]{2004ApJ...617..879L}
{Lin} Y.-T.,  {Mohr} J.~J.,  2004, \mn@doi [\apj] {10.1086/425412}, \href
  {https://ui.adsabs.harvard.edu/abs/2004ApJ...617..879L} {617, 879}

\bibitem[\protect\citeauthoryear{{Mann} \& {Ebeling}}{{Mann} \&
  {Ebeling}}{2012}]{2012MNRAS.420.2120M}
{Mann} A.~W.,  {Ebeling} H.,  2012, \mn@doi [\mnras]
  {10.1111/j.1365-2966.2011.20170.x}, \href
  {https://ui.adsabs.harvard.edu/abs/2012MNRAS.420.2120M} {420, 2120}

\bibitem[\protect\citeauthoryear{{Markevitch} et~al.,}{{Markevitch}
  et~al.}{2000}]{2000ApJ...541..542M}
{Markevitch} M.,  et~al., 2000, \mn@doi [\apj] {10.1086/309470}, \href
  {https://ui.adsabs.harvard.edu/abs/2000ApJ...541..542M} {541, 542}

\bibitem[\protect\citeauthoryear{{Masters} \& {Capak}}{{Masters} \&
  {Capak}}{2011}]{2011PASP..123..638M}
{Masters} D.,  {Capak} P.,  2011, \mn@doi [\pasp] {10.1086/660023}, \href
  {https://ui.adsabs.harvard.edu/abs/2011PASP..123..638M} {123, 638}

\bibitem[\protect\citeauthoryear{{Matthews}, {Morgan}  \& {Schmidt}}{{Matthews}
  et~al.}{1964}]{1964ApJ...140...35M}
{Matthews} T.~A.,  {Morgan} W.~W.,   {Schmidt} M.,  1964, \mn@doi [\apj]
  {10.1086/147890}, \href
  {https://ui.adsabs.harvard.edu/abs/1964ApJ...140...35M} {140, 35}

\bibitem[\protect\citeauthoryear{{McDonald} et~al.,}{{McDonald}
  et~al.}{2012}]{2012Natur.488..349M}
{McDonald} M.,  et~al., 2012, \mn@doi [\nat] {10.1038/nature11379}, \href
  {https://ui.adsabs.harvard.edu/abs/2012Natur.488..349M} {488, 349}

\bibitem[\protect\citeauthoryear{{McDonald}, {Benson}, {Veilleux}, {Bautz}  \&
  {Reichardt}}{{McDonald} et~al.}{2013}]{2013ApJ...765L..37M}
{McDonald} M.,  {Benson} B.,  {Veilleux} S.,  {Bautz} M.~W.,   {Reichardt}
  C.~L.,  2013, \mn@doi [\apjl] {10.1088/2041-8205/765/2/L37}, \href
  {https://ui.adsabs.harvard.edu/abs/2013ApJ...765L..37M} {765, L37}

\bibitem[\protect\citeauthoryear{{McDonald} et~al.,}{{McDonald}
  et~al.}{2014}]{2014ApJ...784...18M}
{McDonald} M.,  et~al., 2014, \mn@doi [\apj] {10.1088/0004-637X/784/1/18},
  \href {https://ui.adsabs.harvard.edu/abs/2014ApJ...784...18M} {784, 18}

\bibitem[\protect\citeauthoryear{{McDonald} et~al.,}{{McDonald}
  et~al.}{2016}]{2016ApJ...817...86M}
{McDonald} M.,  et~al., 2016, \mn@doi [\apj] {10.3847/0004-637X/817/2/86},
  \href {https://ui.adsabs.harvard.edu/abs/2016ApJ...817...86M} {817, 86}

\bibitem[\protect\citeauthoryear{{McDonald} et~al.,}{{McDonald}
  et~al.}{2019}]{2019ApJ...885...63M}
{McDonald} M.,  et~al., 2019, \mn@doi [\apj] {10.3847/1538-4357/ab464c}, \href
  {https://ui.adsabs.harvard.edu/abs/2019ApJ...885...63M} {885, 63}

\bibitem[\protect\citeauthoryear{{McNamara} \& {Nulsen}}{{McNamara} \&
  {Nulsen}}{2012}]{2012NJPh...14e5023M}
{McNamara} B.~R.,  {Nulsen} P.~E.~J.,  2012, \mn@doi [New Journal of Physics]
  {10.1088/1367-2630/14/5/055023}, \href
  {https://ui.adsabs.harvard.edu/abs/2012NJPh...14e5023M} {14, 055023}

\bibitem[\protect\citeauthoryear{{McNamara} \& {O'Connell}}{{McNamara} \&
  {O'Connell}}{1989}]{1989AJ.....98.2018M}
{McNamara} B.~R.,  {O'Connell} R.~W.,  1989, \mn@doi [\aj] {10.1086/115275},
  \href {https://ui.adsabs.harvard.edu/abs/1989AJ.....98.2018M} {98, 2018}

\bibitem[\protect\citeauthoryear{{McNamara} et~al.,}{{McNamara}
  et~al.}{2006}]{2006ApJ...648..164M}
{McNamara} B.~R.,  et~al., 2006, \mn@doi [\apj] {10.1086/505859}, \href
  {https://ui.adsabs.harvard.edu/abs/2006ApJ...648..164M} {648, 164}

\bibitem[\protect\citeauthoryear{{Merritt}}{{Merritt}}{1984}]{1984ApJ...276...26M}
{Merritt} D.,  1984, \mn@doi [\apj] {10.1086/161590}, \href
  {https://ui.adsabs.harvard.edu/abs/1984ApJ...276...26M} {276, 26}

\bibitem[\protect\citeauthoryear{{Newman}, {Treu}, {Ellis}  \& {Sand
  }}{{Newman} et~al.}{2013}]{2013ApJ...765...25N}
{Newman} A.~B.,  {Treu} T.,  {Ellis} R.~S.,   {Sand } D.~J.,  2013, \mn@doi
  [\apj] {10.1088/0004-637X/765/1/25}, \href
  {https://ui.adsabs.harvard.edu/abs/2013ApJ...765...25N} {765, 25}

\bibitem[\protect\citeauthoryear{{Nurgaliev} et~al.,}{{Nurgaliev}
  et~al.}{2017}]{2017ApJ...841....5N}
{Nurgaliev} D.,  et~al., 2017, \mn@doi [\apj] {10.3847/1538-4357/aa6db4}, \href
  {https://ui.adsabs.harvard.edu/abs/2017ApJ...841....5N} {841, 5}

\bibitem[\protect\citeauthoryear{{O'Dea} et~al.,}{{O'Dea}
  et~al.}{2008}]{2008ApJ...681.1035O}
{O'Dea} C.~P.,  et~al., 2008, \mn@doi [\apj] {10.1086/588212}, \href
  {https://ui.adsabs.harvard.edu/abs/2008ApJ...681.1035O} {681, 1035}

\bibitem[\protect\citeauthoryear{{O'Dea} et~al.,}{{O'Dea}
  et~al.}{2010}]{2010ApJ...719.1619O}
{O'Dea} K.~P.,  et~al., 2010, \mn@doi [\apj] {10.1088/0004-637X/719/2/1619},
  \href {https://ui.adsabs.harvard.edu/abs/2010ApJ...719.1619O} {719, 1619}

\bibitem[\protect\citeauthoryear{{Oemler}}{{Oemler}}{1976}]{1976ApJ...209..693O}
{Oemler} A. J.,  1976, \mn@doi [\apj] {10.1086/154769}, \href
  {https://ui.adsabs.harvard.edu/abs/1976ApJ...209..693O} {209, 693}

\bibitem[\protect\citeauthoryear{{Osterbrock}}{{Osterbrock}}{1974}]{1974agn..book.....O}
{Osterbrock} D.~E.,  1974, {Astrophysics of gaseous nebulae}

\bibitem[\protect\citeauthoryear{{Piqueras}, {Conseil}, {Shepherd}, {Bacon},
  {Leclercq}  \& {Richard}}{{Piqueras} et~al.}{2019}]{2019ASPC..521..545P}
{Piqueras} L.,  {Conseil} S.,  {Shepherd} M.,  {Bacon} R.,  {Leclercq} F.,
  {Richard} J.,  2019, in {Molinaro} M.,  {Shortridge} K.,   {Pasian} F.,  eds,
   Astronomical Society of the Pacific Conference Series Vol. 521, Astronomical
  Data Analysis Software and Systems XXVI. p.~545

\bibitem[\protect\citeauthoryear{{Planck Collaboration} et~al.,}{{Planck
  Collaboration} et~al.}{2016}]{2016A&A...594A..27P}
{Planck Collaboration} et~al., 2016, \mn@doi [\aap]
  {10.1051/0004-6361/201525823}, \href
  {https://ui.adsabs.harvard.edu/abs/2016A&A...594A..27P} {594, A27}

\bibitem[\protect\citeauthoryear{{Richard} et~al.,}{{Richard}
  et~al.}{2020}]{2020arXiv200909784R}
{Richard} J.,  et~al., 2020, arXiv e-prints, \href
  {https://ui.adsabs.harvard.edu/abs/2020arXiv200909784R} {p. arXiv:2009.09784}

\bibitem[\protect\citeauthoryear{{Schlafly} \& {Finkbeiner}}{{Schlafly} \&
  {Finkbeiner}}{2011}]{2011ApJ...737..103S}
{Schlafly} E.~F.,  {Finkbeiner} D.~P.,  2011, \mn@doi [\apj]
  {10.1088/0004-637X/737/2/103}, \href
  {https://ui.adsabs.harvard.edu/abs/2011ApJ...737..103S} {737, 103}

\bibitem[\protect\citeauthoryear{{Schneider}, {Gunn}  \& {Hoessel}}{{Schneider}
  et~al.}{1983}]{1983ApJ...268..476S}
{Schneider} D.~P.,  {Gunn} J.~E.,   {Hoessel} J.~G.,  1983, \mn@doi [\apj]
  {10.1086/160973}, \href
  {https://ui.adsabs.harvard.edu/abs/1983ApJ...268..476S} {268, 476}

\bibitem[\protect\citeauthoryear{{Schombert}}{{Schombert}}{1986}]{1986ApJS...60..603S}
{Schombert} J.~M.,  1986, \mn@doi [\apjs] {10.1086/191100}, \href
  {https://ui.adsabs.harvard.edu/abs/1986ApJS...60..603S} {60, 603}

\bibitem[\protect\citeauthoryear{{Shin}, {Clampitt}, {Jain}, {Bernstein},
  {Neil}, {Rozo}  \& {Rykoff}}{{Shin} et~al.}{2018}]{2018MNRAS.475.2421S}
{Shin} T.-h.,  {Clampitt} J.,  {Jain} B.,  {Bernstein} G.,  {Neil} A.,  {Rozo}
  E.,   {Rykoff} E.,  2018, \mn@doi [\mnras] {10.1093/mnras/stx3366}, \href
  {https://ui.adsabs.harvard.edu/abs/2018MNRAS.475.2421S} {475, 2421}

\bibitem[\protect\citeauthoryear{{Sohn}, {Geller}, {Diaferio}  \&
  {Rines}}{{Sohn} et~al.}{2020}]{2020ApJ...891..129S}
{Sohn} J.,  {Geller} M.~J.,  {Diaferio} A.,   {Rines} K.~J.,  2020, \mn@doi
  [\apj] {10.3847/1538-4357/ab6e6a}, \href
  {https://ui.adsabs.harvard.edu/abs/2020ApJ...891..129S} {891, 129}

\bibitem[\protect\citeauthoryear{{Thuan} \& {Romanishin}}{{Thuan} \&
  {Romanishin}}{1981}]{1981ApJ...248..439T}
{Thuan} T.~X.,  {Romanishin} W.,  1981, \mn@doi [\apj] {10.1086/159169}, \href
  {https://ui.adsabs.harvard.edu/abs/1981ApJ...248..439T} {248, 439}

\bibitem[\protect\citeauthoryear{{Tonry}}{{Tonry}}{1985}]{1985AJ.....90.2431T}
{Tonry} J.~L.,  1985, \mn@doi [\aj] {10.1086/113948}, \href
  {https://ui.adsabs.harvard.edu/abs/1985AJ.....90.2431T} {90, 2431}

\bibitem[\protect\citeauthoryear{{Trudeau} et~al.,}{{Trudeau}
  et~al.}{2019}]{2019MNRAS.487.1210T}
{Trudeau} A.,  et~al., 2019, \mn@doi [\mnras] {10.1093/mnras/stz1364}, \href
  {https://ui.adsabs.harvard.edu/abs/2019MNRAS.487.1210T} {487, 1210}

\bibitem[\protect\citeauthoryear{{Voges} et~al.,}{{Voges}
  et~al.}{1999}]{1999A&A...349..389V}
{Voges} W.,  et~al., 1999, \aap, \href
  {https://ui.adsabs.harvard.edu/abs/1999A&A...349..389V} {349, 389}

\bibitem[\protect\citeauthoryear{{Weilbacher} et~al.,}{{Weilbacher}
  et~al.}{2020}]{2020A&A...641A..28W}
{Weilbacher} P.~M.,  et~al., 2020, \mn@doi [\aap]
  {10.1051/0004-6361/202037855}, \href
  {https://ui.adsabs.harvard.edu/abs/2020A&A...641A..28W} {641, A28}

\bibitem[\protect\citeauthoryear{{Wen} \& {Han}}{{Wen} \&
  {Han}}{2013}]{2013MNRAS.436..275W}
{Wen} Z.~L.,  {Han} J.~L.,  2013, \mn@doi [\mnras] {10.1093/mnras/stt1581},
  \href {https://ui.adsabs.harvard.edu/abs/2013MNRAS.436..275W} {436, 275}

\makeatother
\end{thebibliography}
\bibliographystyle{mnras}
\bsp	% typesetting comment

\appendix

\section{Spectroscopic redshifts}

We list in Table~\ref{tab:galz} all galaxies for which spectroscopic redshifts were measured by us with the MUSE and DEIMOS spectrographs (see Section~\ref{sec:obs-spec} for details).

\begin{table}
    \centering
    \begin{tabular}{cccc}
    R.A.\ (J2000) Dec.\ & $z$ & R.A.\ (J2000) Dec & $z$ \\[1mm] \hline\\[-2mm]
02 52 13.86  \,$-$20 58 56.6 & 0.7381 & 02 52 27.03  \,$-$21 00 56.2 & 0.7031 \\
02 52 14.05  \,$-$20 58 06.6 & 0.4874 & 02 52 27.06  \,$-$21 00 52.9 & 0.7029 \\
02 52 14.75  \,$-$20 59 19.6 & 0.6623 & 02 52 27.17  \,$-$21 00 57.0 & 0.7083 \\
02 52 15.20  \,$-$20 59 43.1 & 0.5856 & 02 52 27.22  \,$-$21 00 52.0 & 0.7022 \\
02 52 15.25  \,$-$21 00 20.9 & 0.4876 & 02 52 27.32  \,$-$21 00 43.9 & 0.7056 \\
02 52 15.69  \,$-$20 58 42.8 & 0.7636 & 02 52 27.32  \,$-$21 00 43.9 & 0.7056 \\
02 52 16.63  \,$-$21 03 05.3 & 0.7038 & 02 52 27.34  \,$-$21 00 47.8 & 0.7141 \\
02 52 19.22  \,$-$21 01 59.8 & 0.6738 & 02 52 27.51  \,$-$21 00 49.0 & 0.7026 \\
02 52 20.20  \,$-$21 01 48.4 & 0.7033 & 02 52 27.56  \,$-$21 01 15.4 & 1.0175 \\
02 52 20.53  \,$-$21 01 43.3 & 0.7007 & 02 52 27.65  \,$-$21 00 39.7 & 0.7011 \\
02 52 21.31  \,$-$21 02 08.3 & 0.6706 & 02 52 27.70  \,$-$21 00 48.4 & 0.7009 \\
02 52 22.11  \,$-$21 01 06.2 & 0.4929 & 02 52 27.70  \,$-$21 00 29.8 & 0.5974 \\
02 52 22.71  \,$-$21 01 55.8 & 0.6712 & 02 52 27.71  \,$-$21 01 15.2 & 0.9333 \\
02 52 23.26  \,$-$21 00 20.0 & 0.7100 & 02 52 27.79  \,$-$21 00 21.1 & 1.0190 \\
02 52 23.36  \,$-$21 01 43.2 & 0.6792 & 02 52 28.00  \,$-$21 00 28.3 & 0.5907 \\
02 52 24.94  \,$-$21 00 46.2 & 0.6990 & 02 52 28.16  \,$-$21 00 31.6 & 0.7009 \\
02 52 25.30  \,$-$21 01 03.6 & 1.0200 & 02 52 28.23  \,$-$21 00 52.4 & 0.5970 \\
02 52 25.31  \,$-$21 00 14.0 & 0.7051 & 02 52 28.28  \,$-$21 00 51.2 & 0.7073 \\
02 52 25.38  \,$-$21 01 07.9 & 1.1284 & 02 52 28.35  \,$-$21 00 25.4 & 0.7095 \\
02 52 25.48  \,$-$21 00 52.4 & 1.0162 & 02 52 28.37  \,$-$21 00 40.8 & 0.6925 \\
02 52 25.51  \,$-$21 01 00.0 & 1.0169 & 02 52 28.44  \,$-$21 01 07.4 & 0.7002 \\
02 52 25.86  \,$-$21 00 43.0 & 0.7012 & 02 52 29.03  \,$-$20 59 26.5 & 0.6968 \\
02 52 25.95  \,$-$21 00 23.8 & 0.0000 & 02 52 29.05  \,$-$21 00 39.4 & 0.7025 \\
02 52 26.05  \,$-$21 01 17.5 & 0.4918 & 02 52 29.17  \,$-$21 00 38.3 & 0.7024 \\
02 52 26.08  \,$-$21 00 50.4 & 1.0154 & 02 52 29.21  \,$-$21 00 26.5 & 0.7010 \\
02 52 26.10  \,$-$21 01 10.5 & 1.1288 & 02 52 29.40  \,$-$21 01 12.3 & 0.4855 \\
02 52 26.23  \,$-$21 00 55.7 & 0.6988 & 02 52 29.52  \,$-$21 01 16.6 & 1.1349 \\                   
02 52 26.24  \,$-$21 00 57.8 & 0.6989 & 02 52 29.54  \,$-$20 57 57.3 & 0.6963 \\
02 52 26.29  \,$-$21 00 24.9 & 1.0143 & 02 52 29.60  \,$-$21 01 17.5 & 1.1348 \\
02 52 26.46  \,$-$21 01 16.9 & 0.7124 & 02 52 29.69  \,$-$21 00 48.9 & 0.7041 \\
02 52 26.47  \,$-$21 01 02.6 & 0.7007 & 02 52 30.29  \,$-$21 00 24.2 & 0.6954 \\
02 52 26.59  \,$-$21 00 45.9 & 0.7022 & 02 52 30.34  \,$-$21 01 42.9 & 0.7077 \\
02 52 26.60  \,$-$21 00 47.5 & 0.6966 & 02 52 30.40  \,$-$21 00 47.7 & 0.4886 \\
02 52 26.61  \,$-$21 00 58.1 & 0.7031 & 02 52 30.75  \,$-$21 02 45.1 & 0.7082 \\
02 52 26.62  \,$-$21 00 42.3 & 0.4623 & 02 52 31.21  \,$-$21 01 26.6 & 0.7063 \\
02 52 26.65  \,$-$21 00 22.5 & 0.4364 & 02 52 31.43  \,$-$21 00 43.5 & 0.7058 \\
02 52 26.66  \,$-$21 01 04.3 & 0.5975 & 02 52 32.69  \,$-$21 00 07.5 & 0.7015 \\
02 52 26.67  \,$-$21 01 21.0 & 0.4927 & 02 52 33.53  \,$-$21 03 05.9 & 0.5914 \\
02 52 26.71  \,$-$21 00 52.6 & 0.6882 & 02 52 35.14  \,$-$21 00 30.9 & 0.6926 \\
02 52 26.74  \,$-$21 00 58.6 & 0.7126 & 02 52 35.32  \,$-$21 01 33.2 & 0.7036 \\
02 52 26.78  \,$-$21 00 36.3 & 0.7062 & 02 52 36.76  \,$-$20 58 16.9 & 0.7011 \\
02 52 26.84  \,$-$21 01 12.7 & 0.7106 & 02 52 38.68  \,$-$20 59 11.6 & 0.6910 \\
02 52 26.91  \,$-$21 00 32.6 & 1.4546 & 02 52 38.94  \,$-$20 58 09.6 & 0.2889 \\
02 52 26.94  \,$-$21 00 49.6 & 0.7037 & & \\[1mm] \hline
    \end{tabular}
    \caption{Equatorial coordinates and spectroscopic redshifts of galaxies in the field of eMACS\,J0252 as determined by us from MUSE and DEIMOS observations.}
    \label{tab:galz}
\end{table}

\label{lastpage}
\end{document}